\documentclass[prd,amsmath,amssymb,superscriptaddress,floatfix,nofootinbib,10pt]{revtex4}
\usepackage{times}
\usepackage{amssymb,amsbsy,amsmath,amsfonts}
\usepackage{graphicx}
\usepackage{float}
\usepackage{color}
\usepackage{orcidlink}
\usepackage{morefloats}
\usepackage{rotating}
\usepackage{srcltx}
\usepackage{slashed}
\usepackage{multirow}
\usepackage{verbatim}
\usepackage{hyperref}
\usepackage{tabularx}
\usepackage{bm}
\allowdisplaybreaks[4]

\usepackage{bbding}
\usepackage{threeparttable}

\usepackage{mathrsfs}

\newcommand{\PreserveBackslash}[1]{\let\temp=\\#1\let\\=\temp}
\newcolumntype{C}[1]{>{\PreserveBackslash\centering}p{#1}}
\newcolumntype{R}[1]{>{\PreserveBackslash\raggedleft}p{#1}}
\newcolumntype{L}[1]{>{\PreserveBackslash\raggedright}p{#1}}

\begin{document}

\title{$a_0(980)$ and $f_0(980)$ excitation in the $D^+ \to \pi^+ \eta \eta $ decay}

\author{Jing Song\,\orcidlink{0000-0003-3789-7504}}
\email[]{Song-Jing@buaa.edu.cn}
\affiliation{Center for Theoretical Physics, School of Physics and Optoelectronic Engineering, Hainan University, Haikou 570228, China}
\affiliation{School of Physics, Beihang University, Beijing, 102206, China}%
\affiliation{Departamento de Física Teórica and IFIC, Centro Mixto Universidad de Valencia-CSIC Institutos de Investigación de Paterna, 46071 Valencia, Spain}

\author{Yi-Yao Li\,\orcidlink{0009-0001-6943-4646}}
\email[]{liyiyao@m.scnu.edu.cn}
\affiliation{
State Key Laboratory of Nuclear Physics and 
Technology, Institute of Quantum Matter, South China Normal 
University, Guangzhou 510006, China}
\affiliation{Key Laboratory of Atomic and Subatomic Structure and Quantum Control (MOE), Guangdong-Hong Kong Joint Laboratory of Quantum Matter, Guangzhou 510006, China }
\affiliation{ Guangdong Basic Research Center of Excellence for Structure and Fundamental Interactions of Matter, Guangdong Provincial Key Laboratory of Nuclear Science, Guangzhou 510006, China  }
\affiliation{Departamento de Física Teórica and IFIC, Centro Mixto Universidad de Valencia-CSIC Institutos de Investigación de Paterna, 46071 Valencia, Spain}

\author{Melahat Bayar\, \orcidlink{0000-0002-5914-0126 }}
\email[]{melahat.bayar@kocaeli.edu.tr}
\affiliation{Department of Physics, Kocaeli University, 41380, Izmit, Turkey}
\affiliation{Departamento de Física Teórica and IFIC, Centro Mixto Universidad de Valencia-CSIC Institutos de Investigación de Paterna, 46071 Valencia, Spain}

\author{ Eulogio Oset\,\orcidlink{ https://orcid.org/0000-0002-4462-7919}}
\email[]{oset@ific.uv.es}
\affiliation{Departamento de Física Teórica and IFIC, Centro Mixto Universidad de Valencia-CSIC Institutos de Investigación de Paterna, 46071 Valencia, Spain}
\affiliation{Department of Physics, Guangxi Normal University, Guilin 541004, China}


\begin{abstract}
We have made a thorough study of the $D^+ \to \pi^+ \eta \eta $ reaction, recently measured by the BESIII collaboration, which shows an abnormal strength at high invariant masses in the $\pi\eta$ mass distribution.  We studied in detail the triangle mechanism and the $f_0(1370)$ excitation modes that have been suggested to explain this abnormal feature, and concluded that they are too small to have any important role in the solution to that problem. We have also studied other possible solutions evaluating the contribution of excitations of other $f_0$, $a_0$ and $f_2$ resonances and reached the same conclusion. Unexpectedly, the solution to the problem is found considering the $f_0(980)$ excitation, with the $f_0(980)$ decaying to two $\eta$, which is tied to the $a_0(980)$ production, and well under control. At the same time, the consideration of the $f_0(980)$ excitation solves another non reported problem, which is the $\eta \eta$ mass distribution that comes when only the $a_0(980)$ resonance is allowed to be excited, which produces a large deficiency at low invariant masses compared with experiment. 

\end{abstract}

\maketitle

\section{introduction}\label{intro}

\begin{figure}[H]
  \centering
  \includegraphics[width=0.35\textwidth]{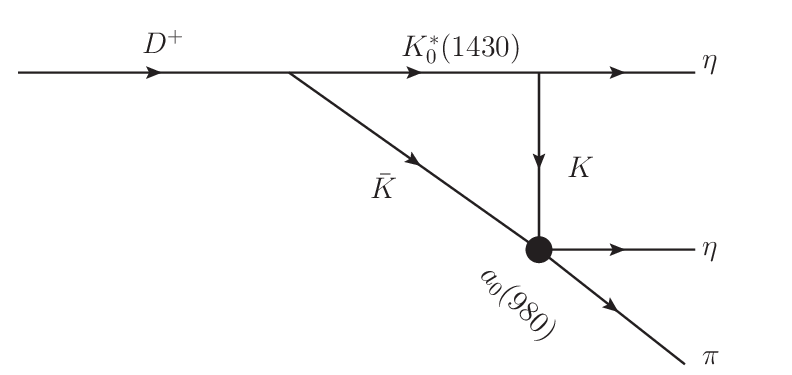}
  \caption{Triangle mechanism suggested in Ref.~\cite{BESIII:2025yag}. }
  \label{fig:1frombesexp}
\end{figure}

The BESIII Collaboration has recently measured the $D^+ \rightarrow \pi ^+ \eta\eta$ reaction~\cite{BESIII:2025yag} and observed a clear signal for the $a_0(980)$  in the $\pi^+ \eta$  mass distribution, however, with an abnormal shape, with a large contribution in the $\pi^+ \eta$ mass region above the $K\bar{K}$ threshold. They attributed this shape to the presence  of a loop rescattering effect (see Fig.~\ref{fig:1frombesexp}). The mechanism is the same that sometime leads to a triangle singularity (TS), although no claims of a TS were made in~\cite{BESIII:2025yag}.

A test for a TS can be easily done by application of Eq.~(18) of~\cite{Bayar:2016ftu}, and one can indeed see that for the nominal mass of the $K_0^{*}(1430)$ the diagram of Fig.~\ref{fig:1frombesexp} does not develop a TS. Among other things because the mass of $K_0^{*}(1430)$ plus the kaon is about 1923 MeV, bigger than the mass of $D^+$~(1869 MeV). However, the $K_0^{*}(1430)$ has a large width, $\Gamma_{K_0^{*}}=270 \pm 80 $ MeV~\cite{ParticleDataGroup:2024cfk}. If we take $\widetilde{M}_{K_0^{*}} = M_{K_0^{*}} - 0.4\Gamma_{{K}^{*}_0} \simeq 1322 $ MeV, then we can see that Eq.~(18) of~\cite{Bayar:2016ftu} is fulfilled and we have a TS for this mass. This means that from a fraction of the mass distribution of the $K_0^{*}(1430)$ one does actually get a TS  from the mechanism of Fig.~\ref{fig:1frombesexp}, which could substantiate the claims of ~\cite{BESIII:2025yag}.

Shortly after the work of~\cite{BESIII:2025yag} was released, a theoretical paper appeared~\cite{Lyu:2025oow} where, based on the nominal mass of the   $K_0^{*}(1430)$, the TS was dismissed, and instead the shape of the $\pi^+ \eta$ mass distribution in~\cite{BESIII:2025yag} was attributed to the contribution of the $f_0(1370)$ state.

The purpose of the present paper is to examine these ideas and present a quantitative contribution from these sources based on empirical information of  different decays. Our first conclusion is that both contributions are small, of the order of 1\% of the experimental branching ratio of the $D^+ \rightarrow \pi ^+ \eta\eta$ reaction, hence, cannot be responsible for the observed shape of the $\pi^+\eta$ mass distribution. 

In view of these results, we have also studied the contribution from $f_0(1500)$, $f_2(1525)$, $a_0(1450)$, and concluded that they also provide too small contributions to explain the experimental shape. Surprisingly, we  observed that linked to the $a_0(980)$ production, there is also excitation  of the $f_0(980)$, decaying to $\eta \eta$, with a strength tied to that of the $a_0(980)$, such that there is no freedom in the relative weights of the two excitation mechanisms. We find that this mechanism naturally explains the large observed strength in the region of $\pi^+ \eta$ mass above the $K\bar{K}$ threshold, and also provides the contribution missing in the $\eta\eta$ mass distribution at low invariant masses.   

\section{formalism}\label{form}

The reaction $D^+ \rightarrow \pi ^+ \eta\eta$  , together with the $D^+ \rightarrow \pi ^+\pi ^0\eta$   one, were studied from the theoretical point of view in Ref.~\cite{Ikeno:2021kzf}. The  process is single Cabibbo suppressed and many diagrams contribute to the reaction. Some approximations were done, but the work predicted clear signals of $a_0(980)$ excitations prior to the BESIII experiment~\cite{BESIII:2025yag}, which confirmed the expectations. We take the complete set of Feynman diagrams for the weak decay at the quark level described in~\cite{Ikeno:2021kzf} and implement the hadronization plus  final state interaction in detail, with the aim of reproducing the experimental spectrum of the BESIII measurements.

\begin{figure}[H]
  \centering
  \includegraphics[width=0.3\textwidth]{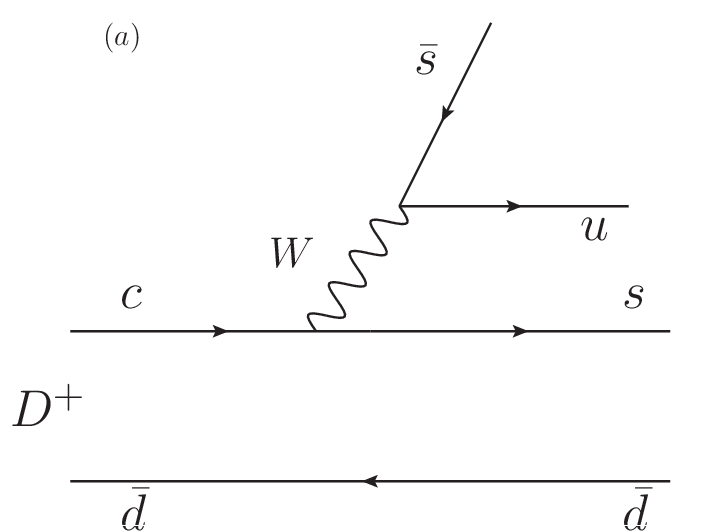}
  \includegraphics[width=0.3\textwidth]{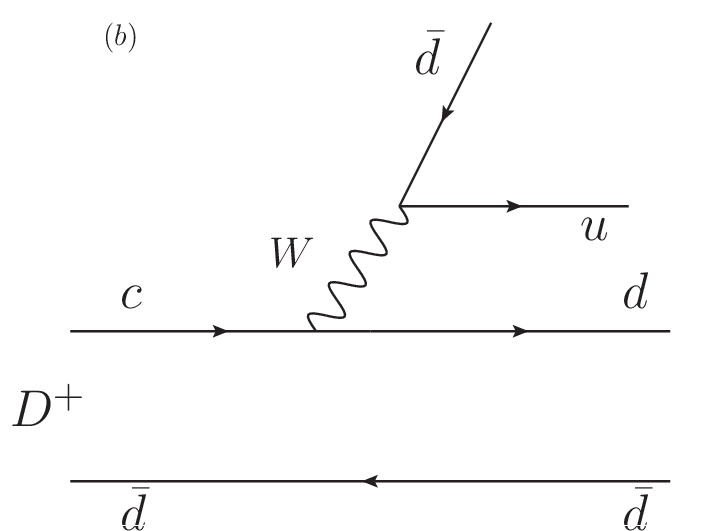}
  \caption{ Diagrams of external emission at the quark level: (a) Cabibbo suppressed $W u \bar s$ vertex, (b) Cabibbo suppressed $Wcd$ vertex. }
  \label{fig:figs1frompaper}
\end{figure}

\begin{figure}[H]
  \centering
  \includegraphics[width=0.3\textwidth]{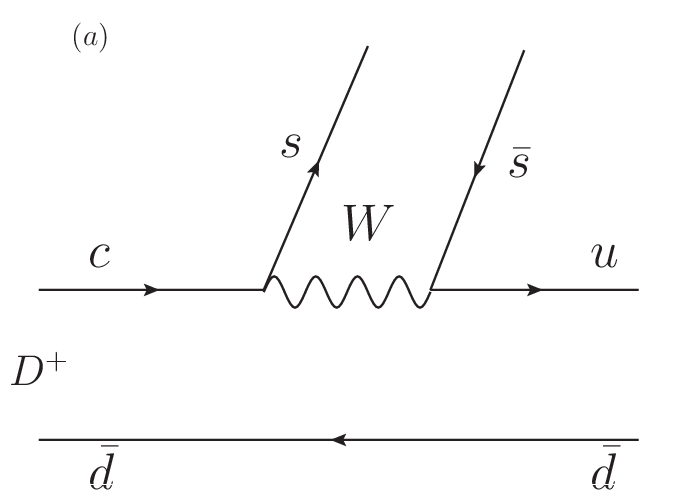}
  \includegraphics[width=0.3\textwidth]{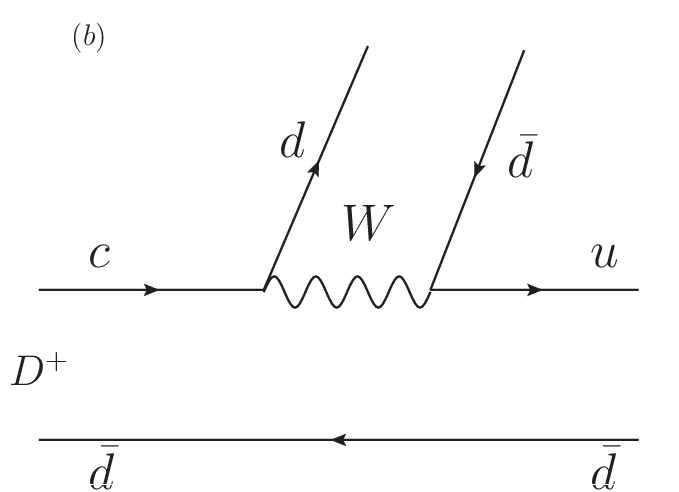}
  \caption{ Diagrams of internal emission at the quark level: (a) Cabibbo suppressed $W u \bar s$ vertex, (b) Cabibbo suppressed $Wcd$ vertex. }
  \label{fig:figs2frompaper}
\end{figure}

\begin{figure}[H]
  \centering
  \includegraphics[width=0.3\textwidth]{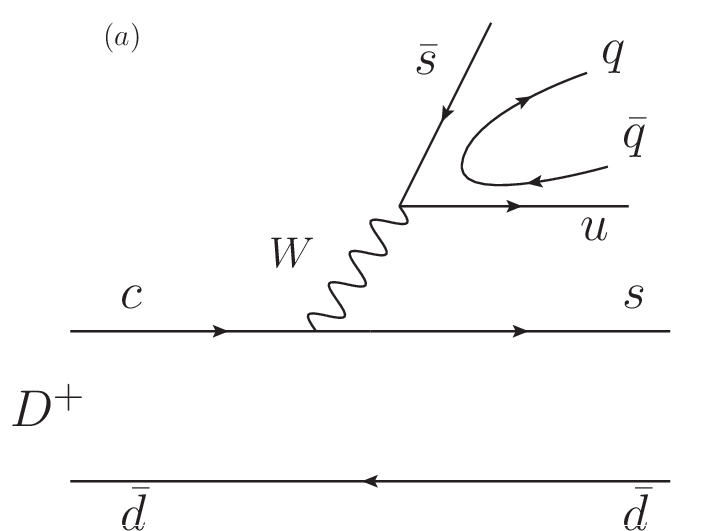}
  \includegraphics[width=0.3\textwidth]{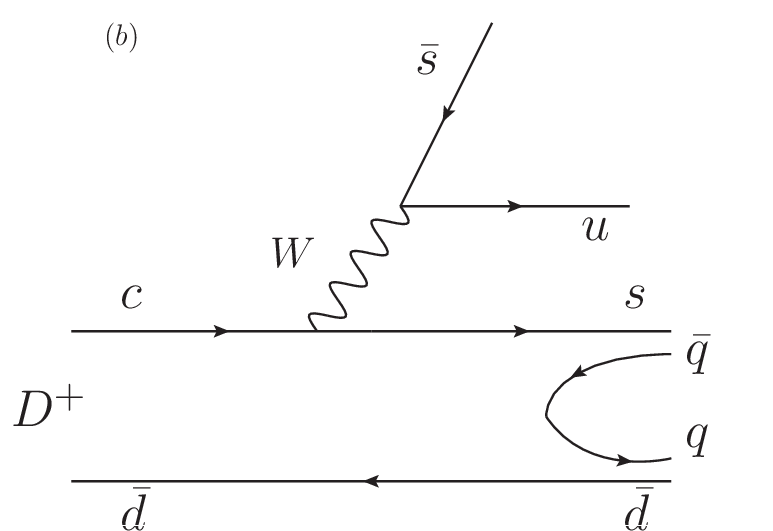}\\
  \includegraphics[width=0.3\textwidth]{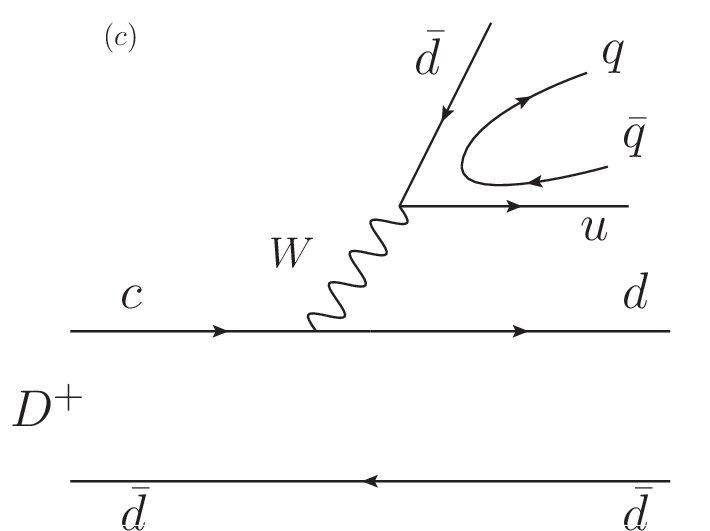}
  \includegraphics[width=0.3\textwidth]{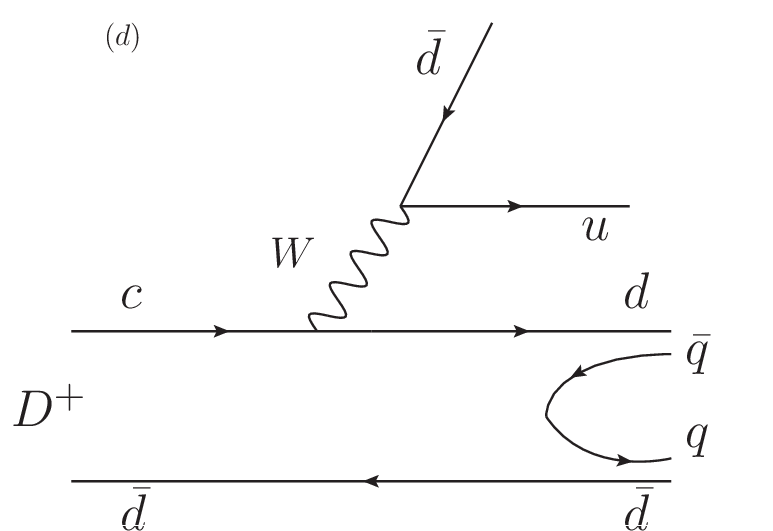}
  \caption{ Hadronization in the diagrams of Fig.~\ref{fig:figs1frompaper}. }
  \label{fig:figs3frompaper}
\end{figure}

\begin{figure}[H]
  \centering
  \includegraphics[width=0.3\textwidth]{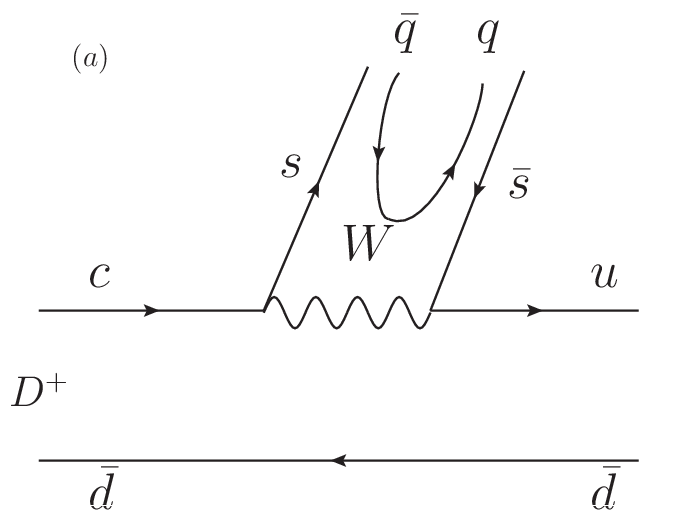}
  \includegraphics[width=0.3\textwidth]{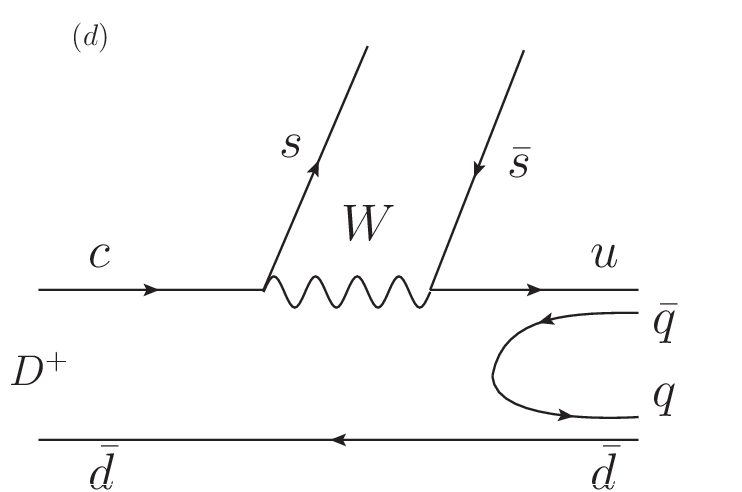}\\
  \includegraphics[width=0.3\textwidth]{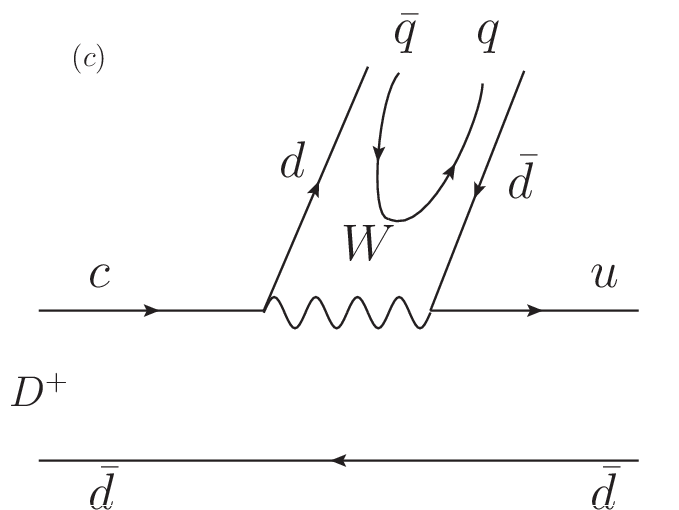}
  \includegraphics[width=0.3\textwidth]{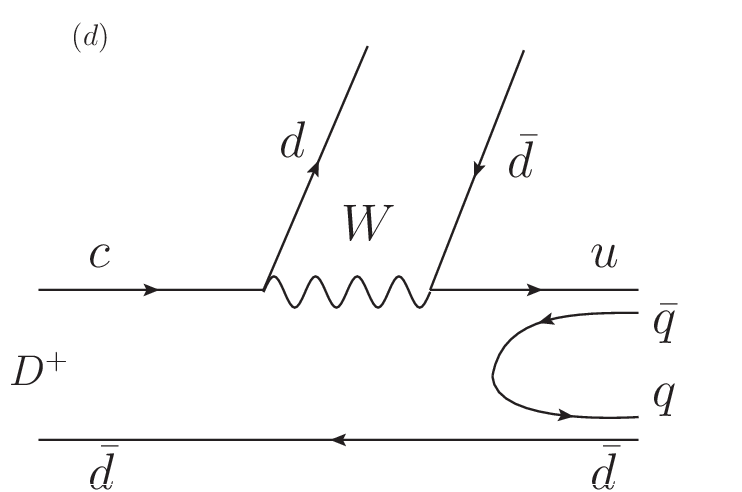}
  \caption{ Hadronization in the diagrams of Fig.~\ref{fig:figs2frompaper}. }
  \label{fig:figs4frompaper}
\end{figure}

The mechanisms considered in~\cite{Ikeno:2021kzf} are shown in Figs.~\ref{fig:figs1frompaper},~~\ref{fig:figs2frompaper} at the quark level. Fig.~\ref{fig:figs1frompaper} shows the two diagrams involved in external emission and Fig.~\ref{fig:figs2frompaper}   those involved in internal emission. In Figs.~\ref{fig:figs3frompaper},~~\ref{fig:figs4frompaper} we take these mechanisms and implement the hadronization of a pair of quarks to produce two mesons.

The hadronization proceeds as follows. Take for instance the diagram of Fig.~\ref{fig:figs3frompaper}~(a). Then, 
\begin{align}\label{eq:hadro}
    u \bar s \to \sum_i u\bar q_iq_i\bar s = P_{1i}P_{i3}\equiv (P^2)_{13},
\end{align}
where $P$ is the $q_i\bar q_j$  matrix, which we write in terms of physical meson states in Eq.~(\ref{eq:hadro})
\begin{equation}\label{eq:pmatr}
  P \equiv 
\left(  
\begin{array}{ccc}
\frac{\pi^0}{\sqrt{2}}+\frac{\eta}{\sqrt{3}}
 & \pi^+ & K^+ \\
\pi^- & -\frac{\pi^0}{\sqrt{2}}+\frac{\eta}{\sqrt{3}}  & K^0  \\
K^- & \bar{K}^0 & -\frac{\eta}{\sqrt{3}}
\end{array}    
\right)\,,
 \end{equation} 
where we have used the standard $\eta-\eta'$ mixing of Ref.~\cite{Bramon:1994cb} and omitted the $\eta'$ which does not play a role here. Similarly to  Eq.~(\ref{eq:hadro}), we can perform the hadronization for all the other diagrams with the following results
\begin{eqnarray} \label{eq:fig3paperFSI}
& & \text{Fig.~\ref{fig:figs3frompaper}~(a)}:~~  (P^2)_{13}\bar{K}^0= \left[\left( \frac{\pi^0}{\sqrt{2}} + \frac{\eta}{\sqrt{3}} \right) K^+ + \pi^+ K^0 - \frac{ K^+\eta}{\sqrt{3}} \right]\bar{K}^0, \\
& & \text{Fig.~\ref{fig:figs3frompaper}~(c)}:~~  (P^2)_{12} \left( -\frac{\pi^0}{\sqrt{2}} + \frac{\eta}{\sqrt{3}} \right) 
= \left[ \left( \frac{\pi^0}{\sqrt{2}} + \frac{\eta}{\sqrt{3}} \right)\pi^++ \pi^+\left( -\frac{\pi^0}{\sqrt{2}} + \frac{\eta}{\sqrt{3}} \right) + K^+\bar{K}^0\right] \left( -\frac{\pi^0}{\sqrt{2}} + \frac{\eta}{\sqrt{3}} \right).
\end{eqnarray}
In~\cite{Ikeno:2021kzf} the $\frac{\eta}{\sqrt{3}}K^+$  terms cancelled for Fig.~\ref{fig:figs3frompaper}~(a) and $\frac{\pi^0\pi^+}{\sqrt{2}}$ terms cancelled in  Fig.~\ref{fig:figs3frompaper}~(c). We correct this here because the order of the fields matters. Indeed, as can be obtained from~\cite{Gasser:1983yg,Scherer:2002tk,Ren:2015bsa} (see also ~\cite{Sun:2015uva,Song:2025eko}), the $WPP$ Lagrangian goes as
\begin{align}\label{eq:princplecancellation}
    W_\mu\langle [P,\partial_\mu P] \rangle
\end{align}
and the terms $\eta K^+,~(-)K^+\eta$   go as  $(\eta \partial_\mu K^+ - K^+\partial_\mu \eta)$ and $(-)(K^+\partial_\mu\eta - \eta \partial_\mu K^+) $ respectively, and hence do not cancel but add. On the other hand, the $\eta \pi^-$  and $\pi^+ \eta$ terms in Fig.~\ref{fig:figs3frompaper}~(c)     cancel rather than add. This said we have for the sum of Figs.~\ref{fig:figs3frompaper}~(a),~\ref{fig:figs3frompaper}~(c)
\begin{align}\label{eq:fig3acpaperFSI}
    \left( \frac{\pi^0}{\sqrt{2}}K^+ + \frac{2}{\sqrt{3}} \eta K^+  + \pi^+K^0\right) \bar{K}^0- \pi^0\pi^+\pi^0 + \sqrt{\frac{2}{3}} \pi^0\pi^+\eta - K^+ \bar{K}^0\frac{\pi^0}{\sqrt{2}} + K^+ \bar{K}^0 \frac{\eta}{\sqrt{3}}.
\end{align}
The Lagrangian of Eq.~(\ref{eq:princplecancellation}) has also the consequence of suppressing terms where the masses of the two mesons are equal or similar since we have the contribution $p^{\mu}(1)-p^{\mu}(2)$   which is dominated by the $\mu=0$ component and then the difference of energies vanishes as an average. There is one term in Eq.~(\ref{eq:fig3acpaperFSI}) which could lead to $\pi^+\eta\eta$ in the final state, the $\eta K^+\bar K^0$ through $K^+\bar K^0\to \pi^+\eta$ rescattering, but given the similar mass of the $\eta$ and $K^+$, we also neglect this term.
In the  $(\pi^+K^0)\bar{K^0}$  we also have $(\omega _{\pi^+} - \omega _{K^0})$   which would also be zero in the SU(3) limit, but given the relatively large momenta of these particles in the phase space, this difference of energies is also small and we will equally neglect this term. As a consequence, we get no contribution from the hadronization of the  $W^\mu  q\bar{q}$ vertex of external emission of figures \ref{fig:figs3frompaper} (a) + \ref{fig:figs3frompaper} (c).
The rest of the terms are the same as in~\cite{Ikeno:2021kzf}.
\begin{eqnarray} \label{eq:fig3bdpaperFSI}
& & \text{Fig.~\ref{fig:figs3frompaper}~(b)}:~~  K^+(P^2)_{32} = K^+ \left[K^-\pi^+ + \bar{K}^0 \left( -\frac{\pi^0}{\sqrt{2}} + \frac{\eta}{\sqrt{3}} \right) - \frac{ \eta}{\sqrt{3}} \bar{K}^0 \right]= K^+\left(K^- \pi^+ - \bar{K}^0  \frac{\pi^0}{\sqrt{2}} \right), \\
& & \text{Fig.~\ref{fig:figs3frompaper}~(d)}:~~  \pi^+(P^2)_{22}  
=  \pi^+ \left( \pi^- \pi^+ + \frac{\pi^0 \pi^0}{2} + \frac{\eta \eta}{3} - \frac{\sqrt{2}}{\sqrt{3}} \pi^0 \eta + K^0 \bar{K}^0 \right). 
\end{eqnarray}

From Fig.~\ref{fig:figs3frompaper}~(d) we get the good term $\frac{\pi^+\eta\eta}{{3}} $. In addition we can have  $K^+K^-\to \eta\eta,~K^0\bar K^0\to \eta\eta$     which has isospin $I=0$ and will be dominated by the $f_0(980)$. 
This term is not considered in~\cite{Lyu:2025oow}. We should also mention a difference with the work of~\cite{Lyu:2025oow}. In Ref.~\cite{Lyu:2025oow} the term $K^+\bar K^0\eta$ was obtained from the intermediate $q\bar q \equiv d\bar d $ in Fig.~\ref{fig:figs3frompaper}~(b). However, as we can see in Eq.~(\ref{eq:fig3bdpaperFSI}), this term is canceled by the contribution from $q\bar q \equiv s\bar s $.
\begin{eqnarray}\label{eq:fig4acpaperFSI} 
& & \text{Fig.~\ref{fig:figs4frompaper}(a)}:~~  (P^2)_{33} \pi^+ = \left( K^- K^+ + \bar{K}^0 K^0 + \frac{\eta \eta}{3} \right) \pi^+, \\
& & \text{Fig.~\ref{fig:figs4frompaper}~(c)}:~~ (P^2)_{22} \pi^+ 
= \left( \pi^- \pi^+ + \frac{\pi^0 \pi^0}{2} + \frac{\eta \eta}{3} - \frac{2}{\sqrt{6}} \pi^0 \eta + K^0 \bar{K}^0 \right) \pi^+.
\end{eqnarray}
We obtain again the term  $\frac{\eta\eta\pi^+}{3}$  in both Fig.~\ref{fig:figs4frompaper}~(a)   and  Fig.~\ref{fig:figs4frompaper}~(c). Once again $K^0\bar{K}^0 \to \eta\eta$   can lead to the final state through the tail of the $f_0(980)$  resonance.
\begin{flushleft}
\begin{equation}\label{eq:fig4bpaperFSI} 
\text{Fig.~\ref{fig:figs4frompaper}~(b)}:~~ \left( -\frac{\eta}{\sqrt{3}} \right) (P^2)_{12}  
= \left( -\frac{\eta}{\sqrt{3}} \right) \left(   \frac{2}{\sqrt{3}} \eta \pi^+  +  K^+ \bar{K}^0 \right),
\end{equation}  

\begin{equation}\label{eq:fig4dpaperFSI} 
 \text{Fig.~\ref{fig:figs4frompaper}~(d)}:~~  
\left( -\frac{\pi^0}{\sqrt{2}} + \frac{\eta}{\sqrt{3}} \right) (P^2)_{12} = \left( -\frac{\pi^0}{\sqrt{2}} + \frac{\eta}{\sqrt{3}} \right) \left( \frac{2}{\sqrt{3}} \eta \pi^+ + K^+ \bar{K}^0 \right).
\end{equation} 
\end{flushleft}
We find that the $\eta\eta\pi^+$ terms, and the $K^+\bar K^0\eta$ terms which could lead to the good final state through $K^+\bar K^0\to \pi^+\eta$, cancel in the sum of Eqs.~(\ref{eq:fig4bpaperFSI}) and ~(\ref{eq:fig4dpaperFSI}). 

We also note that in Ref.~\cite{Lyu:2025oow} the terms from internal emission have not been considered, because they are suppressed with respect to those of external emission. We keep these terms. We shall give a weight $A$ to the terms in external emission and a weight $B$ to those from internal emission, assuming approximately that $B/A\sim \frac{1}{N_c}$, with $N_c=3$, the number of colors. In the final state interaction we shall keep the $\eta\pi^+ \to \eta\pi^+ $   interaction, which  leads to the $a_0(980)$   production, and $K^+K^-\to \eta\eta,~K^0\bar K^0\to \eta\eta,~\pi^+\pi^-\to\eta\eta,~\pi^0\pi^0\to\eta\eta,~\eta\eta\to\eta\eta$ which involve the  $f_0(980)$ tail contribution. 

\begin{figure}[H]
  \centering
  \includegraphics[width=0.23\textwidth]{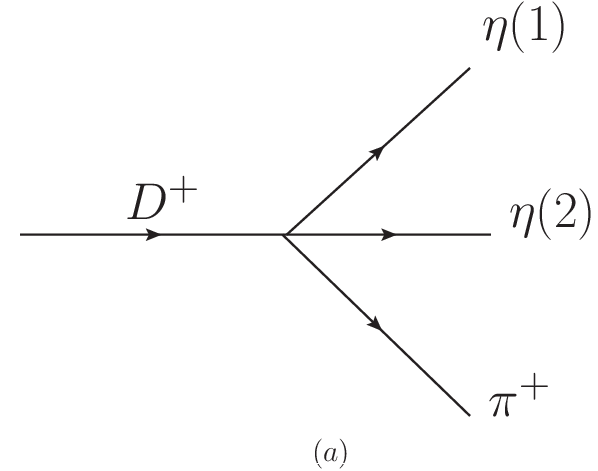}
  \includegraphics[width=0.35\textwidth]{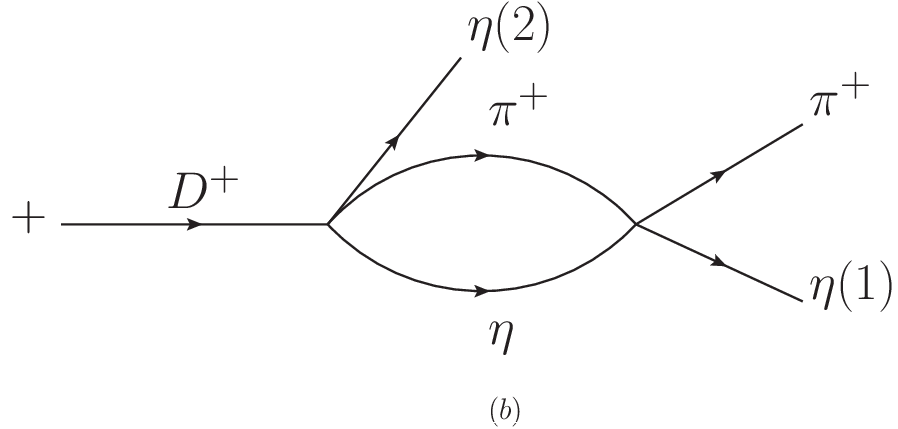}
  \includegraphics[width=0.35\textwidth]{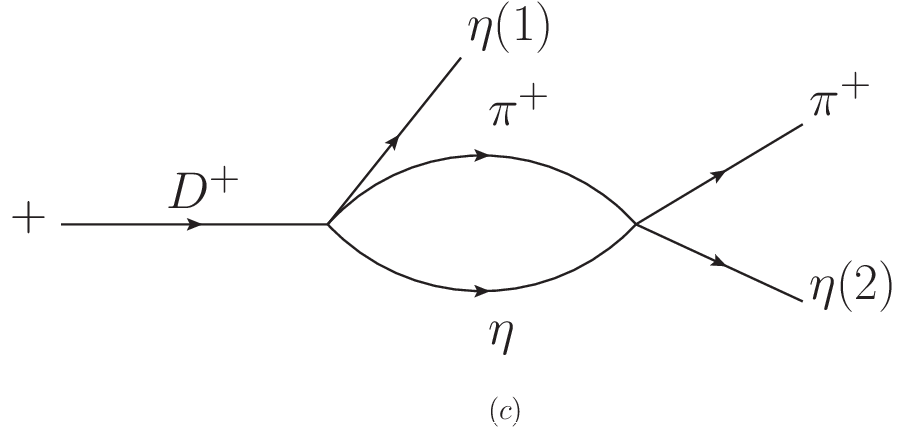}
  \caption{ Tree level and $\pi^+\eta$ rescattering diagrams leading to the $a_0(980)$. }
  \label{fig:RES}
\end{figure}
\begin{figure}[H]
  \centering
  \includegraphics[width=0.3\textwidth]{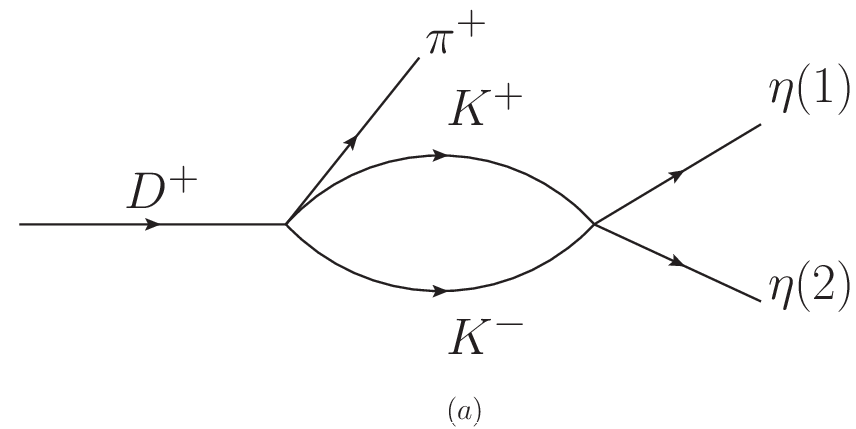}
  \includegraphics[width=0.3\textwidth]{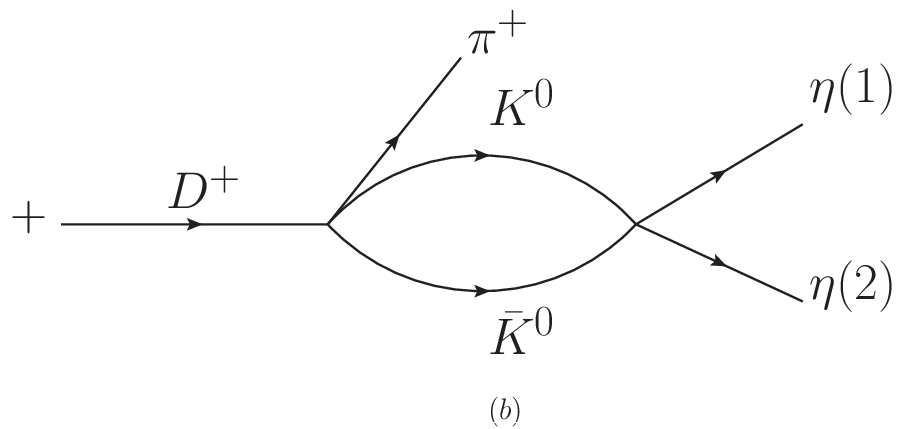}\qquad\qquad\qquad\qquad
  \includegraphics[width=0.3\textwidth]{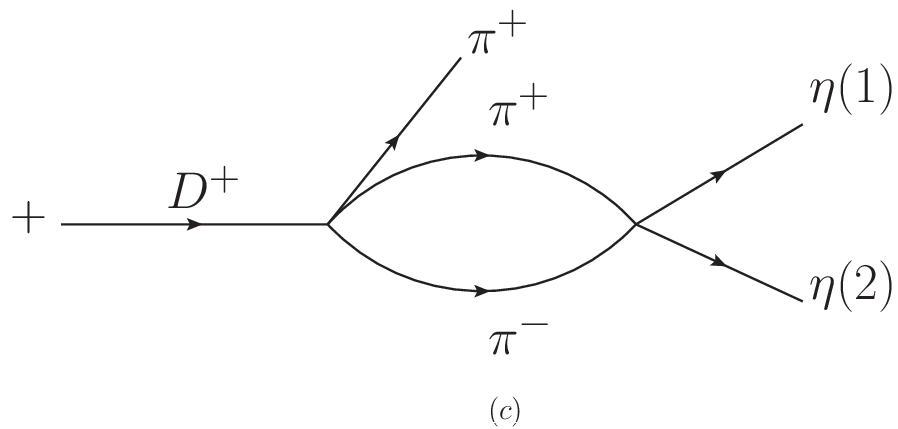}
  \includegraphics[width=0.3\textwidth]{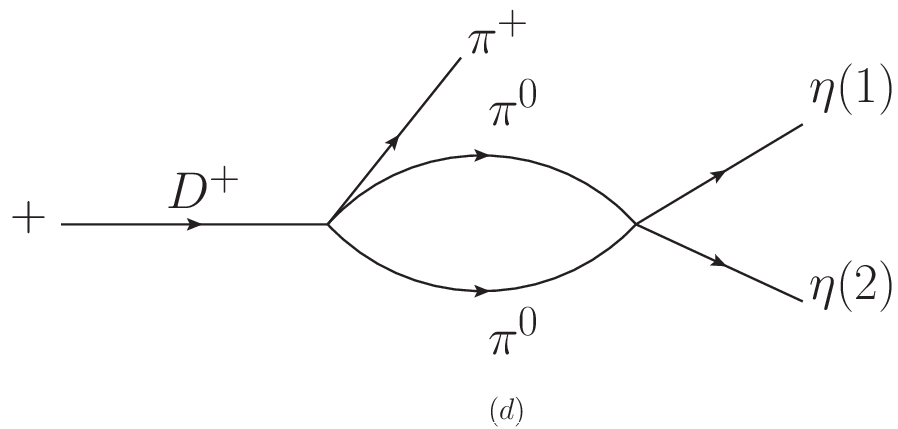}
  \includegraphics[width=0.3\textwidth]{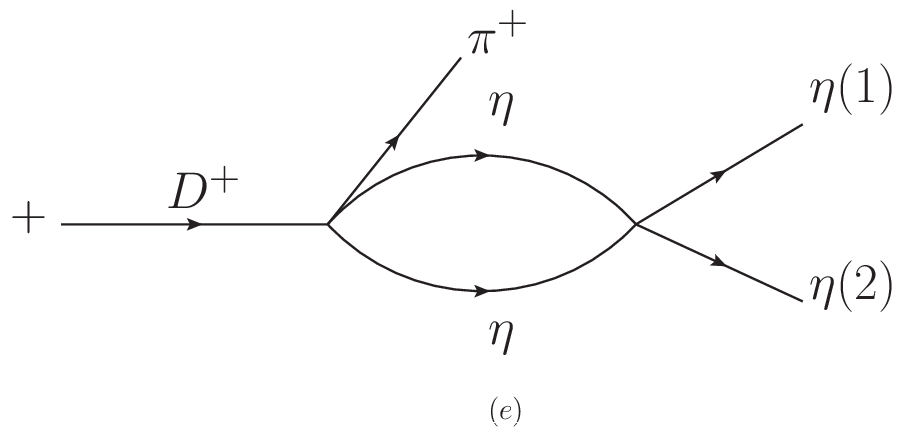}
  \caption{ Rescattering diagrams leading to the $f_0(980)$ production. }
  \label{fig:f980}
\end{figure}

All this said, we have the mechanisms for $D^+ \to \pi^+\eta\eta$ production depicted in Figs.~\ref{fig:RES} and \ref{fig:f980}.

 With the weights $A,~B$ for external, internal emission the valid terms that can lead to the final state are,
\begin{eqnarray}\label{eq:fig34paperFSI} 
& & \text{Fig.~\ref{fig:figs3frompaper}~(b)}:~~   A~K^+K^-\pi^+,\\
& & \text{Fig.~\ref{fig:figs3frompaper}~(d)}:~~   A~\frac{\eta\eta}{{3}}\pi^+;~~A~K^0\bar K^0\pi^+;~~A~\pi^+\pi^-\pi^+;~~A~\pi^+\pi^0\pi^0,\\
& & \text{Fig.~\ref{fig:figs4frompaper}~(a)}:~~   B~\frac{\eta\eta}{{3}}\pi^+;~~B~K^+K^-\pi^+;~~B~K^0\bar K^0\pi^+,\\
& & \text{Fig.~\ref{fig:figs4frompaper}~(c)}:~~   B~\frac{\eta\eta}{{3}}\pi^+;~~B~K^0\bar K^0\pi^+;~~B~\pi^+\pi^-\pi^+;~~B~\pi^+\pi^0\pi^0.
\end{eqnarray} 

With these we write the $D^+$ decay amplitudes corresponding to Figs.~\ref{fig:RES} and \ref{fig:f980}   as 
\begin{align}\label{eq:tRES}
 t_{a_0} =  \frac{2}{{3}}(A+2B)~\Bigg(&1+G_{\pi^+\eta}(M_\text{inv}(\pi^+\eta1))~t_{\pi^+\eta, \pi^+\eta}(M_\text{inv}(\pi^+\eta1))\\\nonumber
& + G_{\pi^+\eta}(M_\text{inv}(\pi^+\eta2))~t_{\pi^+\eta, \pi^+\eta}(M_\text{inv}(\pi^+\eta2)) \Bigg),
 \end{align}
{
the factor 2 in $ t_{a_0}$ for the two identical $\eta$ fields.
 \begin{align}\label{eq:tf980}
 t_{f_0} = & (A+B)~\sqrt{2}\Bigg(G_{K^+K^-}(M_\text{inv}(\eta1\eta2))~t_{K^+K^-,\eta\eta}(M_\text{inv}(\eta1\eta2)) \Bigg)\\\nonumber
 &+ (A+2B)~\sqrt{2}\Bigg(G_{K^0\bar K^0}(M_\text{inv}(\eta1\eta2))~t_{K^0\bar K^0,\eta\eta}(M_\text{inv}(\eta1\eta2)) \Bigg)\\\nonumber
 &+ (A+B)~\Bigg(2G_{\pi^+\pi^-}(M_\text{inv}(\eta1\eta2))~t_{\pi^+\pi^-,\eta\eta}(M_\text{inv}(\eta1\eta2)) \sqrt{2}
 + 2\frac{1}{2}\frac{1}{2}G_{\pi^0\pi^0}(M_\text{inv}(\eta1\eta2))~t_{\pi^0\pi^0,\eta\eta}(M_\text{inv}(\eta1\eta2)) 2\Bigg)\\\nonumber
 &+ \frac{2}{3} (A+2B)~\frac{1}{2}\Bigg(G_{\eta\eta}(M_\text{inv}(\eta1\eta2))~t_{\eta\eta,\eta\eta}(M_\text{inv}(\eta1\eta2)) 2 \Bigg),
 \end{align}
}%
where the factor $\sqrt{2}$ is introduced because in the chiral unitary approach where the $t_{ij}$ scattering matrices are evaluated, the states with identical particles are normalized with the unitary normalization, in this case $\frac{1}{\sqrt{2}}\eta\eta$ (also $\frac{1}{\sqrt{2}}\pi^0\pi^0$), and then the good normalization has to be restored at the end.
For the same reason the factor 2 is implemented to $t_{\pi^0\pi^0,\eta\eta}$ and $t_{\eta\eta,\eta\eta}$ and a factor $\frac{1}{2}$ in the ${\pi^0\pi^0,\eta\eta}$   loops is considered due to the identify of the two particles.
We take all the amplitudes needed here from the work of~\cite{Lin:2021isc}, corresponding to the $\eta-\eta'$ mixing case (Eqs. (A.4) of~\cite{Lin:2021isc}).

The $t_{a_0}$ and $t_{f_0}$ contributions depend on different invariant masses. We evaluate the total decay width using the PDG formula~\cite{ParticleDataGroup:2024cfk} 
\begin{align}
    \frac{d\Gamma}{dM_{12}M_{23}} = \frac{1}{2}\frac{1}{(2\pi)^3}\frac{1}{32m_{D^+}^3}|t_{a_0}+t_{f_0}|^2\times2M_{12}2M_{23},
\end{align}
taking the particles $\eta(1),~\eta(2),~\pi^+(3)$, and the relationship
\begin{align}\label{massrelation}
    M_{12}^2+M_{13}^2+M_{23}^2 = m_{D^+}^2 + m_{\pi^+}^2 + 2m_{\eta}^2,
\end{align}
where the factor $\frac{1}{2}$ comes from having two identical $\eta\eta$ particles at the end. By integrating over one invariant mass with the limits of the PDG ~\cite{ParticleDataGroup:2024cfk} and making permutation of the indices 1,2,3 in the formulas, one can obtain all the invariant mass distributions.

\section{Triangle singularity contribution}
\begin{figure}[H]
  \centering
  \includegraphics[width=0.35\textwidth]{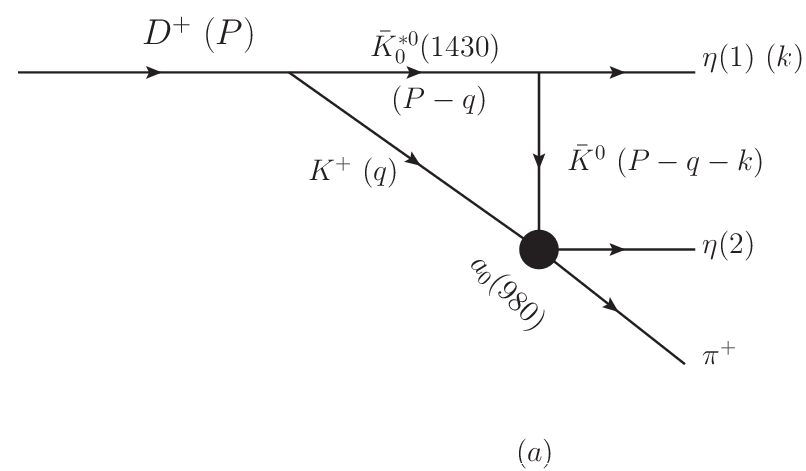}
\includegraphics[width=0.35\textwidth]{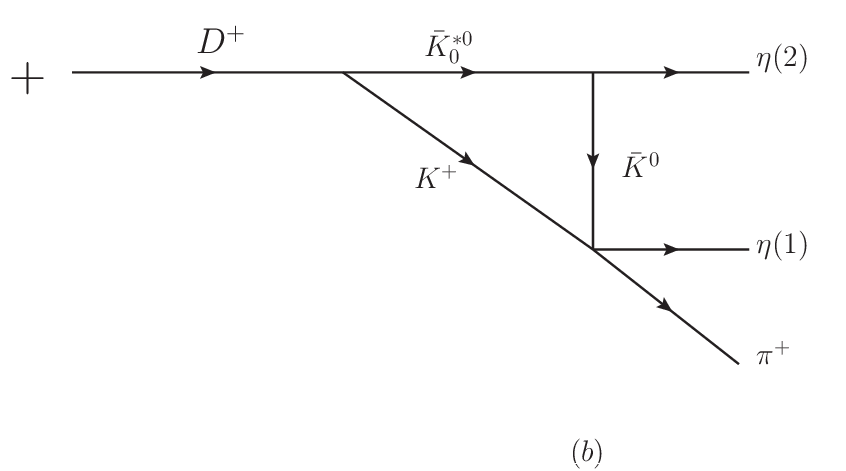}
\includegraphics[width=0.35\textwidth]{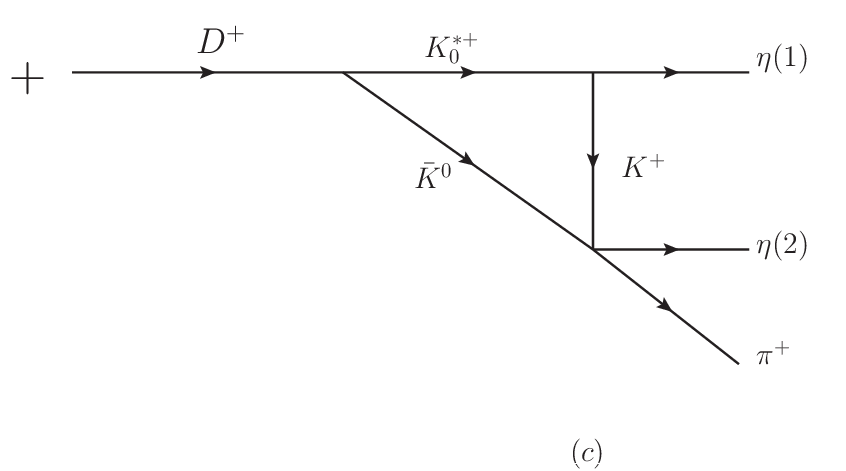}
\includegraphics[width=0.35\textwidth]{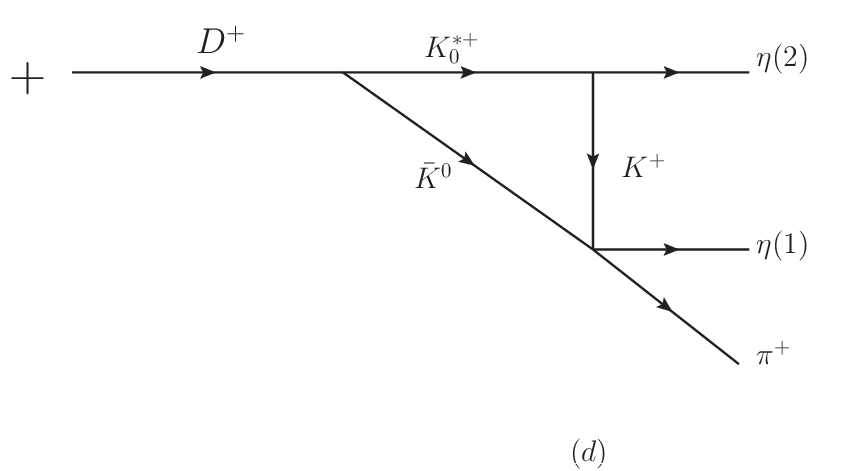}
  \caption{ Triangle diagrams through intermediate 
  $\bar K_0^{*0} K^+$   or $ K_0^{*+} \bar K^0$   production. The momenta of the particles are shown in parentheses. }
  \label{fig:TS_FSI}
\end{figure}

Following the suggestion of the triangle mechanism of the BESIII work~\cite{BESIII:2025yag} we look at the contribution of diagrams in Fig.~\ref{fig:TS_FSI}
The evaluation of the decay rate from the triangle mechanism of Fig.~\ref{fig:TS_FSI} can be done providing not just a shape but also absolute numbers, using empirical information available in the literature. The $\bar K^{0}_0K^+\to \pi^+\eta$   scattering amplitude is available from the chiral unitary approach~\cite{Oller:1997ti}. We need the   $D^+\to\bar K^{*0}_0(1430)K^+ $  coupling and the one for $\bar K^{*0}_0(1430)\to \bar K^0\eta$    which we obtain from decay widths reported in the literature.

\subsection{$D^+\to\bar K^{*}_0 (1430) K$ and $ K^{*}_0\to  K\eta$  }
In the PDG~\cite{ParticleDataGroup:2024cfk} we find the branching ratio
\begin{equation}\label{eq:BR_DpK1430}
\text{BR}[D^+ \to \bar K^{*0}_0(1430)K^+;~\bar K^{*0}\to K^-\pi^+] = (1.82\pm0.35)\times10^{-3},
\end{equation}

\begin{equation}\label{eq:BR_K1430Kpi}
\text{BR}[  K^{*}_0(1430) \to K \pi ] = (93\pm10)\%,
\end{equation}

\begin{equation}\label{eq:BR_K1430Keta}
\text{BR}[  K^{*}_0(1430) \to K \eta ] = (8.6^{+2.7}_{-3.4})\%.
\end{equation}
Since $\bar K^{*0}_0$   has $I=1/2,~I_3=1/2$, using Clebsch–Gordan coefficients the relative rate of $\bar K^{*0}_0$ decaying to $\pi^+K^-$ is $\frac{2}{3}$ versus $\frac{1}{3}$  for $\pi^0 \bar K^0$. Thus, considering Eq.~(\ref{eq:BR_DpK1430}) and~(\ref{eq:BR_K1430Kpi}) we find

\begin{equation}\label{eq:BR_DpK1430K+}
\text{BR}[D^+ \to \bar K^{*0}_0(1430)K^+] = \frac{1.82}{\frac{2}{3}\times 0.93} \simeq (3.0\pm0.6)\times10^{-3}.
\end{equation}
By taking the experimental width~\cite{ParticleDataGroup:2024cfk}
\begin{align}
    \Gamma_{K^*_0(1430)} = 270\pm 80~\text{MeV}
\end{align}
and using Eq.~(\ref{eq:BR_K1430Keta}) we find 
\begin{align}\label{eq:kstc0width}
    \Gamma_{\bar K^{*0}_0(1430) \to \bar K^0\eta } =  270 \frac{8.6}{100}~\text{MeV}= 23.2\pm 6.9~\text{MeV}.
\end{align}
The decay width for $D^+\to \bar K^{*0}_0K^+$   is given in terms of the $\bar K^{*0}_0K^+$  coupling, $g_{D^+,\bar K^{*}_0K}$, by 
\begin{align}
    \Gamma_{D^+\to\bar K^{*0}_0K^+} = \frac{1}{8\pi}\frac{1}{m_{D^+}^2}g^2_{D^+,\bar K^{*}_0K}~q
\end{align}
with 
\begin{align}
    q = \frac{\lambda^{1/2}(m_{D^+}^2, m_{\bar K^{*}_0}^2, m_{K^+}^2)}{2m_{D^+}}\Theta(m_{D^+}-m_{\bar K^{*}_0}-m_{K^+}).
\end{align}
The first thing we realize is that for the nominal mass of the  $K^*_0$ this decay cannot occur. However, it is possible thanks to the large width of the $K^{*}_0$   that gives rise to a wide $K^{*}_0$   mass distribution. We can get the decay width through a convolution as
\begin{align}\label{convolution}
 \Gamma_{D^+\to\bar K^{*}_0K} &= \frac{1}{N} \int_{(m_{K^{*}_0} - 2\Gamma_{K^{*}_0 })^2}^{(m_{K^{*}_0} + 2\Gamma_{K^{*}_0})^2} \, d\tilde{m}^2 \left(-\frac{1}{\pi}\right) \times \text{Im} \left[\frac{1}{\tilde{m}^2 - m_{ K^{*}_0}^2 + i~m_{ K^{*}_0} \Gamma_{ K^{*}_0}} \right] \Gamma_{D^+\to\bar K^{*}_0K} ~(m_{ K^{*}_0}\equiv \tilde{m}),
\end{align}
with a normalization factor
\begin{align}
 N &= \int_{(m_{ K^{*}_0} - 2\Gamma_{K^{*}_0})^2}^{(m_{ K^{*}_0} + 2\Gamma_{K^{*}_0})^2} \, d\tilde{m}^2 \left(-\frac{1}{\pi}\right) \text{Im} \left[\frac{1}{\tilde{m}^2 - m_{ K^{*}_0}^2 + i~m_{ K^{*}_0} \Gamma_{ K^{*}_0}} \right].
\end{align}
By equating Eq.~(\ref{convolution}) to the experimental branching ratio of Eq.~(\ref{eq:BR_DpK1430K+}) we obtain
{
\begin{align}
    \frac{(g_{D^+,\bar K^{*}_0K})^2}{\Gamma_{D^+}} = 2584.12~\text{MeV}.    
\end{align}
}

On the other hand, using 
\begin{align}
    \Gamma_{\bar K^{*}_0,~\bar K \eta} = \frac{1}{8\pi}\frac{1}{m_{ K^{*}_0}^2}g^2_{\bar K^{*}_0,\bar K \eta} q_\eta
\end{align}
with 
\begin{align}
    q_\eta = \frac{\lambda^{1/2}(m_{ K^{*}_0}^2, m_{\bar K^{0}}^2, m_{\eta}^2)}{2m_{ K^{*}_0}}
\end{align}
and taking the experimental width of Eq.~(\ref{eq:kstc0width}) we obtain
{
\begin{align}
    ({g_{\bar K^{*}_0, \bar K \eta}})^2 =2.44707\times10^6  ~\text{MeV}^2.    
\end{align}
}

\newpage

\subsection{The triangle mechanism width}
We are now in a position to make a precise evaluation of the rate of $D^+$ decay through the triangle mechanism of Fig.~\ref{fig:TS_FSI}. Next we evaluate the amplitude for the mechanism of Fig.~\ref{fig:TS_FSI}~(a) as
\begin{align}\label{eq:tTS}
-it_{TS} = \int\,& \frac{d^4q}{(2\pi)^4} (-i)\, g_{D^+,\bar K^{*}_0K}\,(-i)\,g_{\bar K^{*}_0, \bar K \eta}\,(-i)\, t_{K^+\bar K^0, \pi^+\eta}\\\nonumber
&\frac{i}{(P-q)^2-m^2_{\bar K^{*}_0}+i\epsilon}\, \frac{i}{q^2-m^2_{K^+}+i\epsilon} \, \frac{i}{(P-q-k)^2-m^2_{\bar K^0}+i\epsilon}.
\end{align}

The evaluation can be extremely simplified if we take into account that the TS appears when all the intermediate particles in the loop are placed on shell~\cite{Landau:1959fi,Bayar:2016ftu}. This means that we can separate the positive and negative energy parts of a meson propagator as
\begin{align}
    \frac{1}{q^2-m^2+i\epsilon} = \frac{1}{2 \omega(q)} \left( \frac{1}{q^0-\omega(q)+i\epsilon}-\frac{1}{q^0+\omega(q)-i\epsilon} \right)
\end{align}
with $\omega(q)=\sqrt{\vec{q}^{~2}+m^2}$, and keep only the positive energy part, which is the only one that can be placed on shell. Then the amplitude of Eq.~(\ref{eq:tTS}) can be written in the $D^+$ rest frame, where  $\vec{P} =0$, as
\begin{equation}\label{eq:tTS2}
\begin{aligned}
{t}_{\mathrm{TS}} = & i~ g_{D^+,\bar K^{*}_0K}\,g_{\bar K^{*}_0, \bar K \eta}\, t_{K^+\bar K^0, \pi^+\eta}(M_\text{inv}( \pi^+\eta)) \int\, \frac{d^4q}{(2\pi)^4}\, \frac{1}{2\omega_{K^{*}_0}(\vec{q}\,)} \frac{1}{2\omega_{K^+}(\vec{q}\,)} \frac{1}{2\omega_{\bar{K}^0}(\vec{q}+\vec{k})} \\
&\, \frac{1}{P^0 - q^0 - \omega_{K^{*}_0}(\vec{q}\,) + i\epsilon}\, \, \frac{1}{q^0 - \omega_{K^+}(\vec{q}\,) + i\epsilon}\, \frac{1}{P^0 - q^0 - k^0 - \omega_{\bar{K}^0}(\vec{q}+\vec{k}) + i\epsilon}\, .
\end{aligned}
\end{equation}
where $\omega_i(p)=\sqrt{m_i^2+\vec{p}^{~2}}$. Next we perform the $q^0$ integration using Cauchy's theorem and obtain
\begin{align}\label{eq:tTS3}
    {t}_{\mathrm{TS}} = & g_{D^+,\bar K^{*}_0K}\,g_{\bar K^{*}_0, \bar K \eta}\, t_{K^+\bar K^0, \pi^+\eta} (M_\text{inv} (\pi^+\eta)) \int\, \frac{d^3q}{(2\pi)^3}\, \, \frac{1}{2\omega_{K^{*}_0}(\vec{q}\,)} \frac{1}{2\omega_{K^+}(\vec{q}\,)} \frac{1}{2\omega_{\bar{K}^0}(\vec{q}+\vec{k})} \\\nonumber
& \frac{1}{P^0 - \omega_{K^+}(\vec{q}\,) - \omega_{K^{*}_0}(\vec{q}\,) + i\frac{\Gamma_{\bar K^*_0}}{2}}\, \, \frac{1}{P^0 - \omega_{K^+}(\vec{q}\,) - k^0 - \omega_{\bar{K}^0}(\vec{q}+\vec{k}) + i\epsilon}\, ,
\end{align}
where we have replaced $\omega_{K^*_0}-i\epsilon$ by $\omega_{K^*_0}-\frac{i \Gamma_{ K^*_0}}{2}$  to account for the $K^*_0$  width. We have also introduced the factor $\Theta(q_\text{max}-|\vec{q}^{~*}|)$, where $q^*$ is the $K^+$ momentum in the rest frame of the $\pi \eta$ forming the $a_0(980)$ resonance, given by
\begin{equation}\label{cmq}
\begin{aligned}
 \vec{q}~^* &= \left[\left(\frac{E_R}{M_{23}} - 1\right)\, \frac{\vec{q} \cdot \vec{k}}{\vec{k}^2} +\, \frac{q^0}{M_{23}}\right] \vec{k} + \vec{q}, 
\end{aligned}
\end{equation}
where $M_{23} \equiv M_{\mathrm{inv}}(\pi^+ \eta2)$, $q_\text{max}$ is the regulator of the loop in the cut off regularization used to obtain the $t_{K\bar{K},\pi \eta}$ amplitudes in~\cite{Lin:2021isc},  $k$ is the momentum of $\eta_1$, $E_R=\sqrt{M_{23}^2+\vec{k}^2}$ and $q^0= \sqrt{m_{K^+}^2+\vec{q}^{~2}}$. The factor $\Theta(q_\text{max}-|\vec{q}^{~*}|)$ appears because the calculated scattering amplitudes factorize as $t(\vec{q},\vec{q}~{'})=t~\Theta(q_\text{max}-|\vec{q}~|)\Theta(q_\text{max}-|\vec{q}{~'}|)$ ~\cite{Gamermann:2009uq}.

The final amplitude properly symmetrized corresponds to the sum of the diagrams of Fig.~\ref{fig:TS_FSI}. First we assume that Fig.~\ref{fig:TS_FSI} (a) and \ref{fig:TS_FSI} (c) give the same contribution, hence we multiply by 2 the $t$ matrix of Eq.~(\ref{eq:tTS}), and then symmetrize the amplitudes,
exchanging $\eta(1)\leftrightarrow \eta(2)$ to account for the diagrams of Figs.~\ref{fig:TS_FSI} (b) and  ~\ref{fig:TS_FSI} (d). The first two amplitudes depend on the invariant mass $M_{23}$ labeling the states $\eta{(1)},~\eta{(2)},~\pi^+{(3)}$. The second pair of diagrams are a function of $M_{13}$. Then we use the formulas of the PDG to obtain the width,
\begin{align}\label{eq:massdis_tot}
    \frac{d \Gamma}{dM_{12} dM_{23}} = \frac{1}{2} \frac{1}{(2\pi)^3}\frac{1}{32m_{D^+}^3}|2t_{\eta_1\eta_2}+2t_{\eta_2\eta_1}|^2 \times 2M_{12}2M_{23},
\end{align}
where the factor $1/2$ comes from having two identical $\eta$ in the final state.

Considering Eq.~(\ref{massrelation}) and integrating Eq.~(\ref{eq:massdis_tot}) over $M_{12}$ and $M_{23}$ with the limits of the~\cite{ParticleDataGroup:2024cfk} we have the total width coming from the triangle mechanism. If we integrate over $M_{23}$ or $M_{12}$, we obtain the corresponding mass distributions.

\subsection{$f_0(1370)$ contribution}
\begin{figure}[H]
  \centering
  \includegraphics[width=0.35\textwidth]{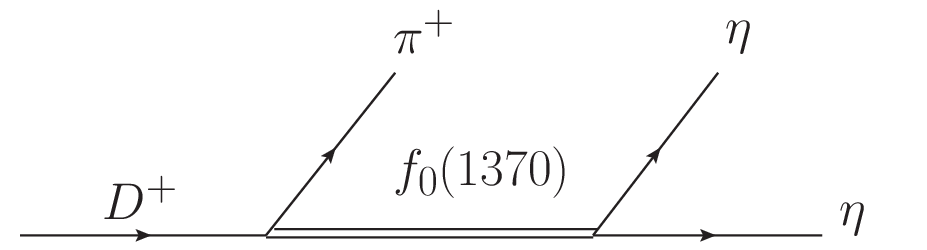}
  \caption{ $D^+\to \pi^+\eta\eta$ decay driven by the $f_0(1370)$ resonance. }
  \label{fig:f1370_FSI}
\end{figure}

In the work of~\cite{Lyu:2025oow}, the contribution of the $f_0(1370)$, according to the diagram of Fig.~\ref{fig:f1370_FSI} is suggested as a way to obtain the higher part of the $\pi^+\eta$ mass spectrum, and its strength is fitted to the experiment.

The $f_0(1370)$ is obtained as a molecular state of $\rho\rho$ in~\cite{Molina:2008jw}, and, as a consequence, it cannot decay into $\eta\eta$. Yet, in the work of~\cite{Geng:2008gx} the approach of~\cite{Molina:2008jw} is extended to SU(3) and the state gets a small component of $K^*\bar{K}^*$ which allows the decay into $\eta\eta$. Actually this decay is reported in the PDG~\cite{ParticleDataGroup:2024cfk} as
\begin{align}\label{widthee}
    \frac{\Gamma(\eta\eta)}{\Gamma(4\pi)} = (28\pm11)\times10^{-3}~(\cite{Anisovich:2001ay}) ;\qquad(4.7\pm2)\times10^{-3}~(\cite{WA102:2000lao}),
\end{align}
and at the same time
\begin{align}
    \text{BR}(D^+ \to \pi^+f_0(1370);~f_0 \to \pi^+ \pi^-)=(8\pm4) \times 10^{-5}.
\end{align}
Taking into account the $f_0 \to \pi^0\pi^0$ decay width which according to isospin should be $1/2$ that of $\pi^+\pi^-$. We have 
\begin{align}
    \text{BR}(D^+ \to \pi^+ f_0(1370);~f_0 \to \pi\pi)=(12\pm6)\times10^{-5}.
\end{align}
The ratio $\Gamma_{\pi\pi}/\Gamma_{\text{tot}}$ in the PDG are contradictory. We then rely on theoretical findings. In~\cite{Geng:2008gx} the $\pi\pi$ decay branching ratio is the largest. One obtains $\Gamma_{\text{tot}}(f_0(1370))\sim 260$ MeV; $\Gamma_{\pi\pi }(f_0(1370))\sim 210$ MeV. Hence one finds a ratio
\begin{align}\label{widthpp}
    \frac{\Gamma_{\pi\pi}}{\Gamma_{\text{tot}}}\simeq 0.81.
\end{align}
And assuming that the width is {exhausted} by the $4\pi$ and $2\pi$ channels, one finds
\begin{align}
    \frac{\Gamma_{4\pi}}{\Gamma_{\pi\pi}}=0.24.
\end{align}
This number is in agreement with the findings of Ref.~\cite{Albaladejo:2008qa} $\Gamma_{4\pi}/\Gamma_{\pi\pi} = 0.30 \pm 0.12$, which is also in line with the interval 0.10-0.25 claimed in the experiment of Ref.~\cite{Bugg:2007ja}. We shall take the result of Eq.~(\ref{widthpp}) for our estimates. Taking into account
all this information we find finally, by taking the largest branching ratio of Eq.~(\ref{widthee})
\begin{align}\label{widthDto1370}
    \text{BR}(D^+ \to \pi^+ f_0;~f_0 \to \eta\eta) \approx \frac{12 \times 10^{-5}}{0.81\times 0.24}\times 28\times 10^{-3}=1.7\times 10^{-5}.
\end{align}
With the other branching ratio of Eq.~(\ref{widthee}), this number would be $2.8\times10^{-6}$. There are large uncertainties in this estimate,  {but we should compare this rate with the experimental branching ratio for $D^+ \to \pi^+\eta\eta$ found in~\cite{BESIII:2025yag}}
\begin{align}\label{BREXP}
    \text{BR}(D^+ \to \pi^+\eta\eta) = (3.67\pm0.12\pm0.06)\times 10^{-3}.
\end{align}
We see that in the best of the cases, the rate in Eq.~(\ref{widthDto1370}) about 200 times smaller than experiment. Thus, we can safely ignore this mechanism. 

\subsection{Contribution from $D^+ \to \eta a_0(1450);~ a_0 \to \pi^+\eta$}

\begin{figure}[H]
  \centering
  \includegraphics[width=0.35\textwidth]{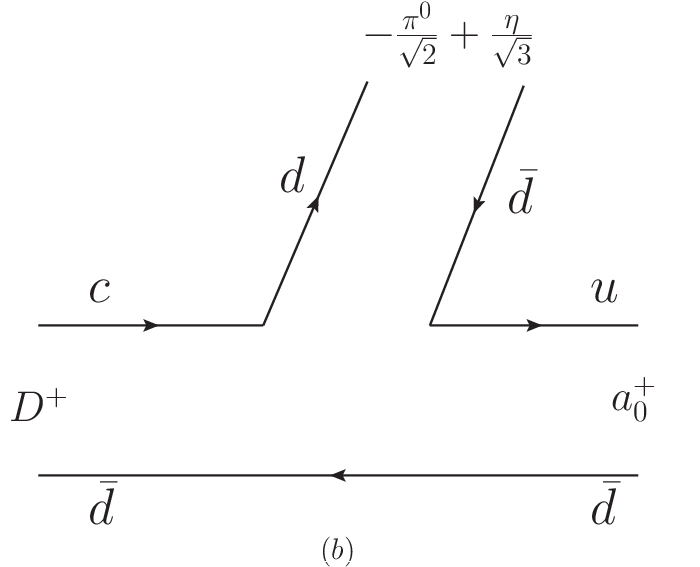}
  \includegraphics[width=0.35\textwidth]{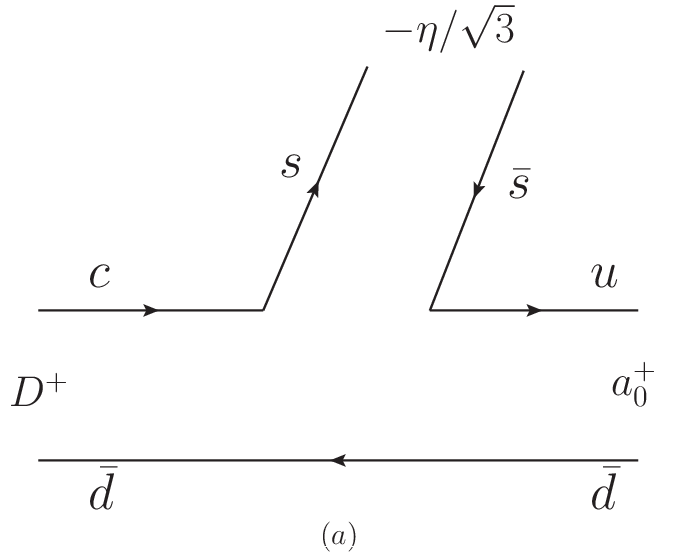}
  \caption{ Diagrams for $D^+\to \pi^0a_0^+,~\eta a_0^+$.
 }
  \label{fig:internala0}
\end{figure}

There is no $D^+ \to \eta a_0(1450)$ reported in the PDG, but there is however the decay mode $D^+ \to \pi^0 a_0(1450)$ with the following branching ratio
\begin{align}
    \text{BR} (D^+ \to a_0(1450) \pi_0; ~a^+ \to \pi^+\eta)=(1.4\pm0.6)\times 10^{-4}.
\end{align}
{We can envisage the production in Fig.~\ref{fig:internala0}}
As we can see, in this Cabibbo suppressed mechanism, the $\pi^0a_0^+$ can be produced, but the $\eta a_0^+$ production cancels with the sum of these diagrams (note that the $a_0(980)$ would be produced through hadronization of the $u \bar d$ pair, but the cancellation occurs anyway).

\subsection{Contribution for $D^+ \to f'_2(1525)\pi^+;f_2'(1525) \to \eta\eta$}

\begin{figure}[H]
  \centering
  \includegraphics[width=0.35\textwidth]{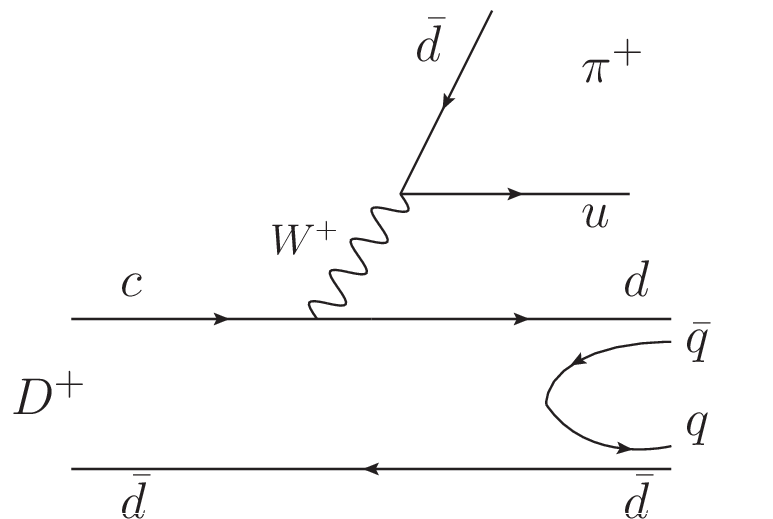}
  \caption{ Mechanisms for $D^+\to \pi^+f_2(1270)$ through hadronization of the $d\bar{d}$  pair.
 }
  \label{fig:external}
\end{figure}

In the PDG, there is no $D^+ \to f'_2(1525)\pi$ reported. However, one finds
\begin{align}
    \text{BR}(D^+ \to f_2(1270) \pi^+; ~f_2 \to \pi^+\pi^-)=(4.58\pm0.28)\times 10^{-4}.
\end{align}
The $f_0(1270)$ is obtained in~\cite{Molina:2008jw} and ~\cite{Geng:2008gx} as a $\rho\rho$ bound state, decaying basically in $\pi\pi$, with a branching fraction of $84\%$. Since $f_2 \to \pi^0\pi^0$ has $1/2$ the strength of $f_2 \to \pi^+ \pi^-$, hence we have 
\begin{align}
    \text{BR}(D^+ \to f_2(1270)\pi^+;~ f_2 \to \pi\pi) = (6.87\pm 0.42)\times 10^{-4},
\end{align}
and considering the $84\%$ branching ratio of $f_2$ decaying to $\pi\pi$ we obtain
\begin{align}
    \text{BR}(D^+ \to f_2(1270)\pi^+)=(8.18 \pm 0.50)\times10^{-4}.
\end{align}
From the molecular picture of~\cite{Geng:2008gx} the $D^+ \to f_2(1270)\pi^+$ can proceed via external emission as shown in Fig.~\ref{fig:external}.

The hadronization of {$d\bar{d}$} in Fig.~\ref{fig:external} to produce two vectors, which will give rise to $f_2(1270)$ is done again as 
\begin{align}
    d\bar{d} \to \sum_i d \bar{q}_i q_i \bar{d}=\sum_i V_{2i}V_{i2}=(V^2)_{22},
\end{align}
where now the $d\bar{q}_i$ becomes a vector meson. The $q_i\bar{q}_j$ matrix written in terms of physical vectors is given by
\begin{align}
    \left(
    \begin{matrix}
        \frac{\rho_0}{\sqrt{2}} + \frac{\omega}{\sqrt{2}}& \rho^+ & K^{*+}\\
        \rho^- & \frac{-\rho_0}{\sqrt{2}} + \frac{\omega}{\sqrt{2}}
        &K^{*0}\\
        K^{*-} & \bar{K}^{*0} & \phi
    \end{matrix}
    \right),
\end{align}
and then 
\begin{align}
    (V^2)_{22} = \rho^-\rho^+ + \left( -\frac{\rho^0}{\sqrt{2}}+\frac{\omega}{\sqrt{2}} \right)^2+K^{*0}\bar{K}^{*0}.
\end{align}
We see that we obtain a $\rho\rho$ component and a $K^{*}\bar{K}^{*}$ with the same strength. In~\cite{Geng:2008gx} the $f_2(1270)$ is mostly $\rho\rho$ and the $f_2'(1525)$ mostly $K^{*}\bar{K}^{*}$. Within the picture of~\cite{Geng:2008gx} for these two particles, we should expect similar results for $D^+ \to \pi^+f_2(1270)$ and $D^+ \to \pi^+ f_2'(1525)$, and here we also assume 
\begin{align}
    \text{BR}(D^+ \to \pi^+ f_2'(1525))\simeq 8\times 10^{-4}.
\end{align}
If now we take from the PDG that 
\begin{align}
    \text{BR}(f'_2(1525)\to\eta\eta) = (10.3\pm2.2)\%,
\end{align}
then
\begin{align}
    \text{BR}(D^+ \to \pi^+ f_2'(1525);~f'_2\to\eta\eta)\simeq 8\times 10^{-5}.
\end{align}
This is bigger than the rate of $f_0(1270)$ production but still a factor $50$ times smaller than the branching ratio of $D^+\to \pi^+\eta\eta$ of Eq.~(\ref{BREXP}). The fact that the  { $f_2'$ decay  requires the  $\eta\eta$ in $D$-wave to conserve angular momentum and parity} also softens the interference with the $S$-wave main mechanisms described so far. We have written the explicit amplitude for this process and added it coherently to  $t_{a_0}$ and have seen indeed that the correction introduced by the new mechanism is indeed negligible.

We have chosen resonances in the region of 1500 MeV because these are the potential candidates giving relevant signals in the Dalitz plot of $D^+ \to \pi^+\eta\eta$. We have looked into $f_2(1270)$ and $f_0(1710)$ resonances, which are both of $VV$ character in~\cite{Geng:2008gx} and seen that they barely contribute in the phase space of the $D^+ \to \pi^+ \eta\eta$ reaction.

After this discussion, in the next section we discuss the results of the triangle singularity and the $a_0(980)$, $f_0(980)$ contributions.

\section{Results}
\subsection{$a_0(980)$ contribution}

\begin{figure}[H]
  \centering
  \includegraphics[width=0.35\textwidth]{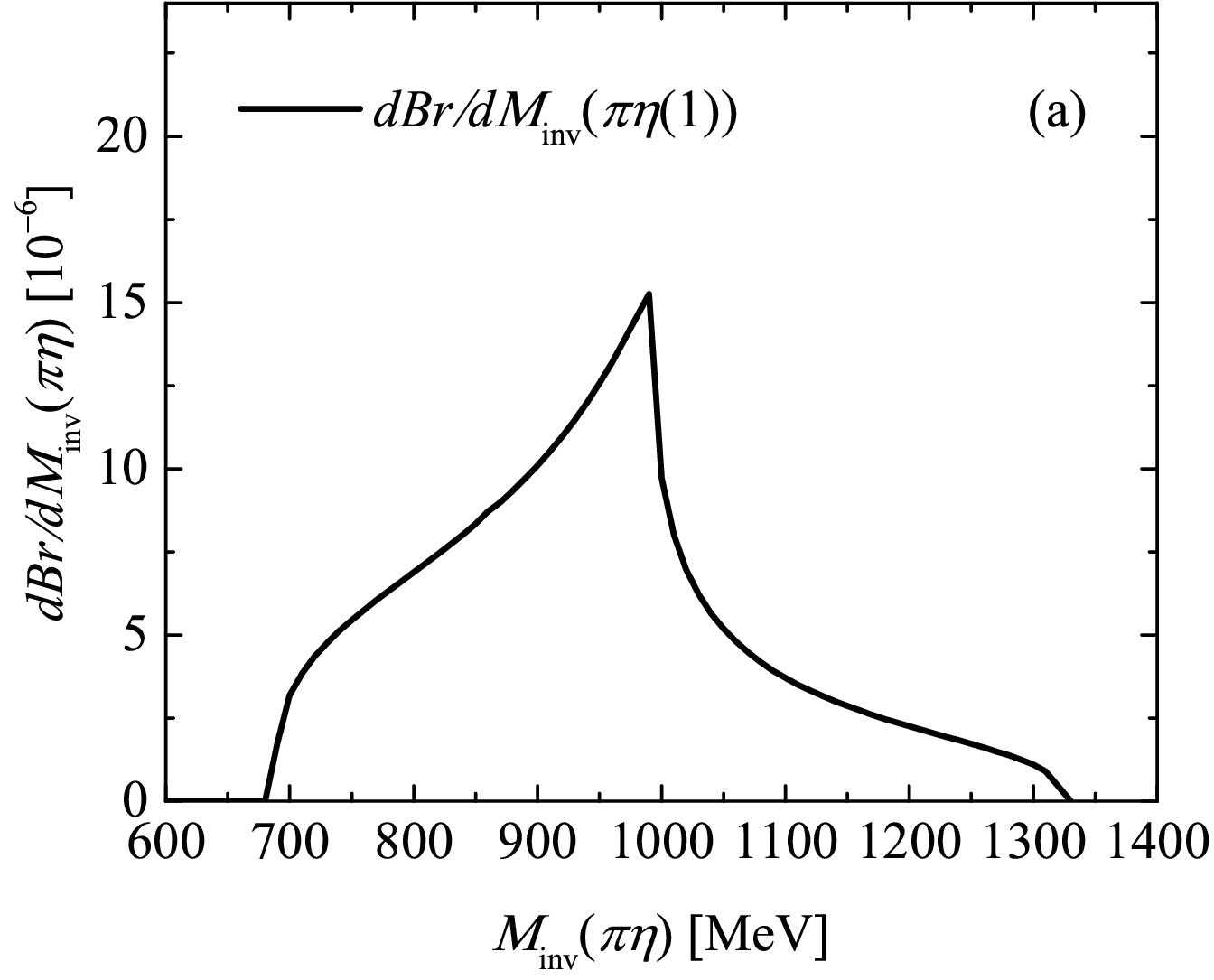 }
  \includegraphics[width=0.35\textwidth]{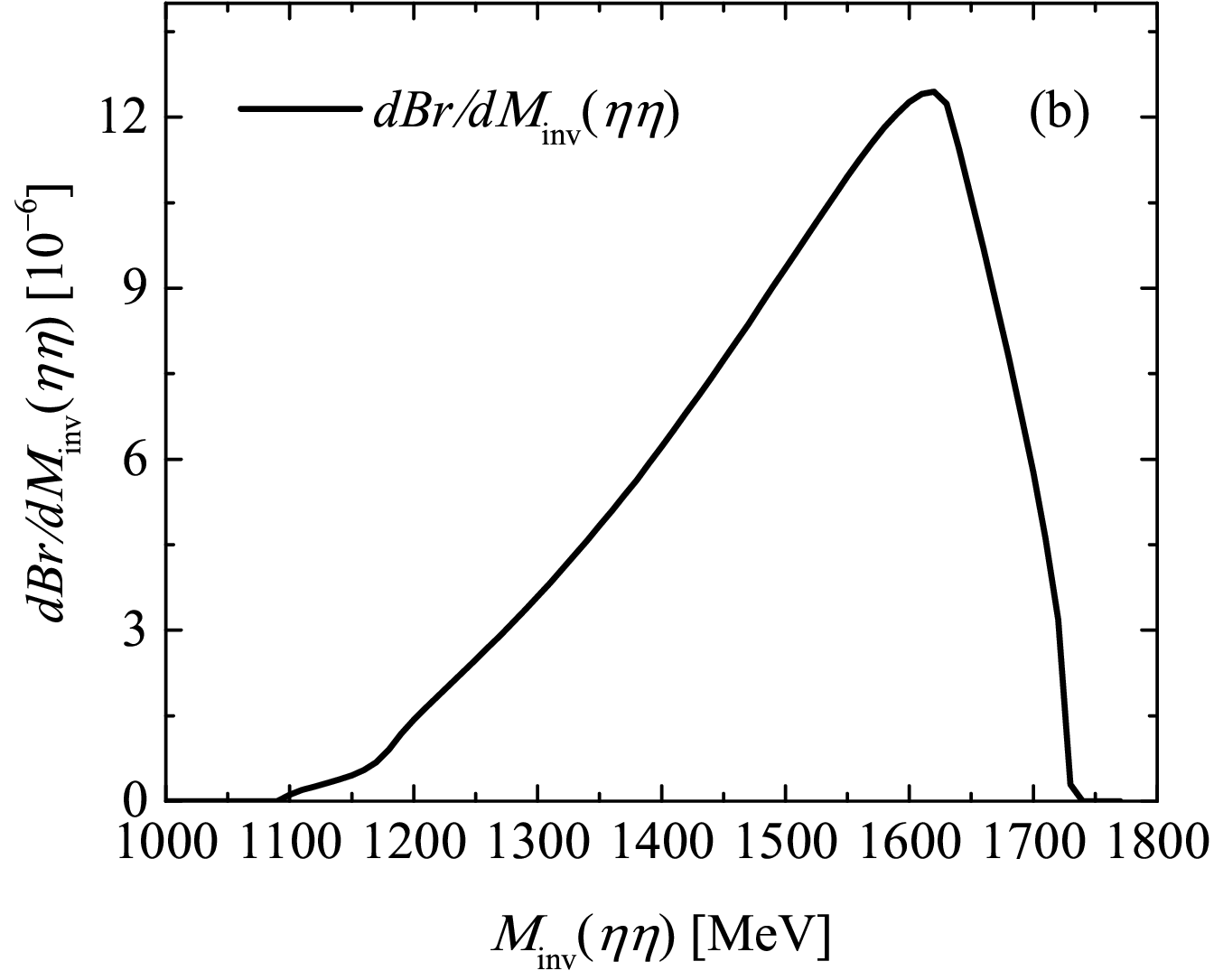 }
  \caption{ (a) $\pi^+\eta$ mass distribution, (b) $\eta\eta$ mass distribution considering only the $a_0(980)$ contribution. }
  \label{fig:RES_massdis}
\end{figure}

In Fig.~\ref{fig:RES_massdis}, we have the results that we obtain for the $D^+ \to \pi^+ \eta \eta$ mass distributions with the $t_{a_0}$ amplitude of Eq.~(\ref{eq:tRES}). We have taken $A=1$ and $B=1/3 $.

We consider the $M_{\pi^+\eta}$ distribution by looking at just one $\eta$, such that the integral over the mass of Figs.~\ref{fig:RES_massdis}(a) and ~\ref{fig:RES_massdis}(b) are the same. When comparing with the data later, we will sum the $\pi^+\eta(1)$ and $\pi^+\eta(2)$ distributions. 
What we observe is that the shape of the two mass distributions are very different than the experimental data shown in {Fig.~\ref{fig:a0+f980_Bplay}}. The shape of $a_0(980)$ mass distribution resembles the shape seen in other experiments, and we note that there is a large discrepancy with the experimental results both at low and particularly at large $M_{\pi^+\eta}$ masses. The other relevant feature in this figure is that the $\eta\eta$ mass distribution has a large peak around 1600 MeV in sheer discrepancy with the data. We shall see later that both problems are solved when we include the contribution of the $f_0(980)$ resonance, $t_{f_0}$ of Eq.~(\ref{eq:tf980}).

\subsection{Triangle mechanism contribution}

We have taken the formula of Eq.~(\ref{eq:massdis_tot}) and  integrated over the two invariant masses the TS amplitude and we obtain the result
\begin{align}
    {\text{BR}(D^+\to \pi^+\eta\eta(TS)) = 6.16 \times 10^{-5}}.
\end{align}
As we can see, the rate obtained, which could be calculated with absolute values, is very small compared to the experimental rate of Eq.~(\ref{BREXP}).

\begin{figure}[H]
  \centering
  \includegraphics[width=0.45\textwidth]{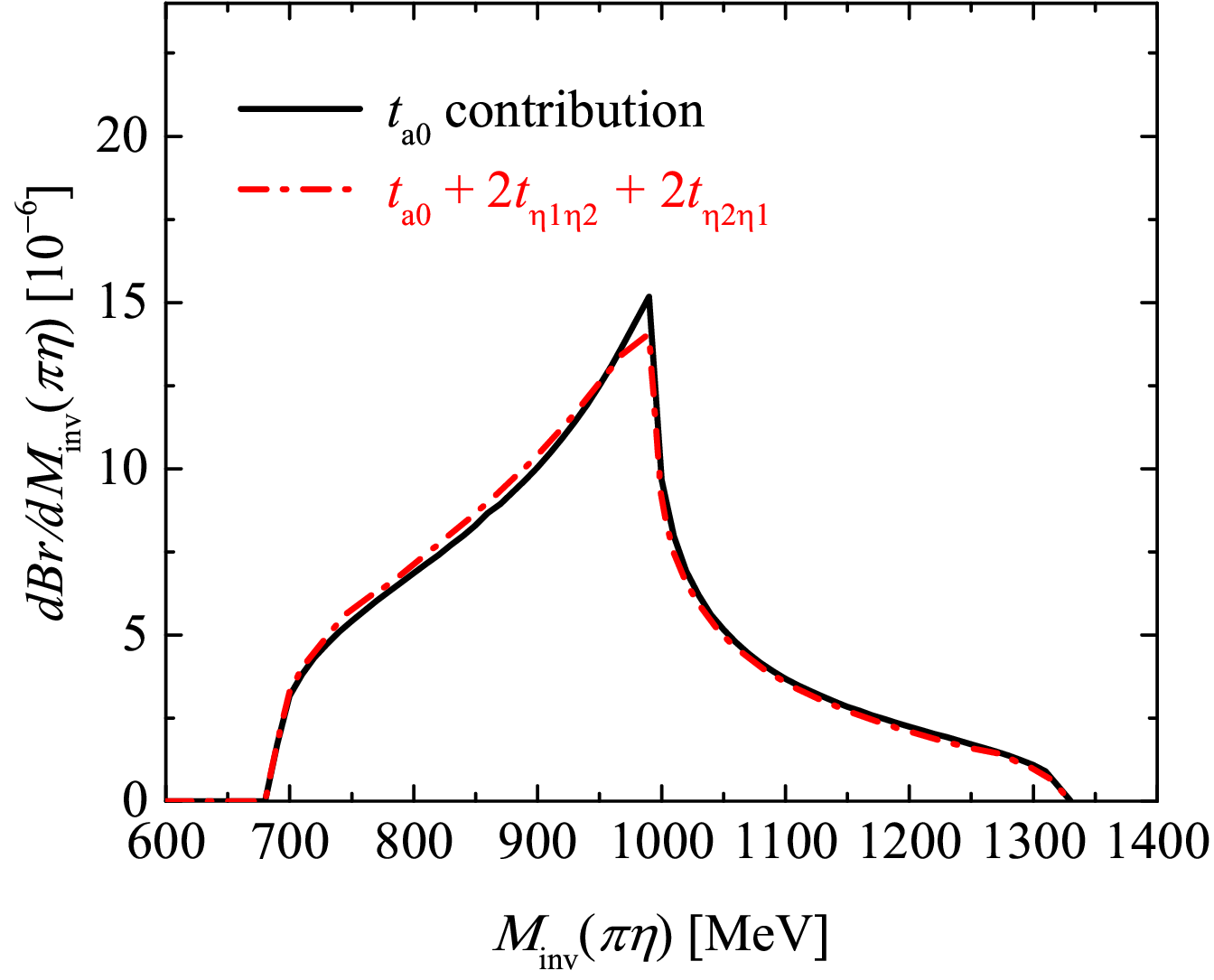 }
  \caption{ Mass distribution of $\pi^+\eta$ with and without the triangle mechanism. }
  \label{fig:TS_massdis}
\end{figure}

We have also looked to see if the coherent sum of amplitudes could give some visible effect. For this we take $A+2B$ of Eq. (\ref{eq:tRES}) such as to get the integrated branching ratio for $D^+ \to \pi^+ \eta\eta$ with the $t_{a_0}$ term alone and then sum the amplitudes  $2t_{\eta_1\eta_2}+2t_{\eta_2\eta_1}$ of Eq.~ (\ref{eq:massdis_tot}) and show the mass distribution, in Fig.~\ref{fig:TS_massdis}.
As we can see, even summing coherently the $a_0(980)$ and the triangle diagram coherently, the effects of the TS mechanisms are negligible.

\section{$f_0(980)$ contribution}

The $f_0$ amplitude comes from the $t_{K\bar K,\eta\eta},~t_{\pi\pi,\eta\eta},~t_{\eta\eta,\eta\eta}$. These will provide a contribution to the $\eta\eta$ mass distribution starting from the $\eta\eta$ threshold. Yet, as one can see in Fig.~\ref{fig:a0+f980_Bplay}, the experimental $\eta\eta$ mass distribution stretches up to 1700 MeV where the chiral unitary approach does not provide good amplitudes. When this happens it is customary to cut the amplitudes and extrapolate them adiabatically, reducing them gradually as $M_\text{inv}(\eta\eta)$ grows up, as a realistic resonance amplitude would do. This is done in~\cite{Debastiani:2016ayp,Toledo:2020zxj,Song:2025ofe} using the formula
\begin{align}\label{eq:cut}
    G(M_{\text{inv}})t_{ij}(M_{\text{inv}}) \to G(M_{\text{cut}})t_{ij}(M_{\text{cut}}) e^{-\alpha(M_{\text{inv}}-M_{\text{cut}})},  M_{\text{inv}}>M_{\text{cut}},
\end{align}
with values of $M_\text{cut}\sim1050-1150$~MeV and $\alpha\sim 0.0040-0.0080~\text{MeV}^{-1}$ are used. We will calculate the mass distributions within this range to evaluate uncertainties of our approach. Since the strength of the $a_0$ and $f_0$ are tied to the coefficients $A$ and $B$, and interfere in the mass distributions, we should make the same cuts to the $a_0$ and $f_0$ amplitudes implementing Eq.~(\ref{eq:cut}) to the $\pi\eta$ and $\eta\eta$ invariant masses in the amplitudes, and this is done in the subsequent calculations.

\begin{figure}[H]
  \centering
  \includegraphics[width=0.35\textwidth]{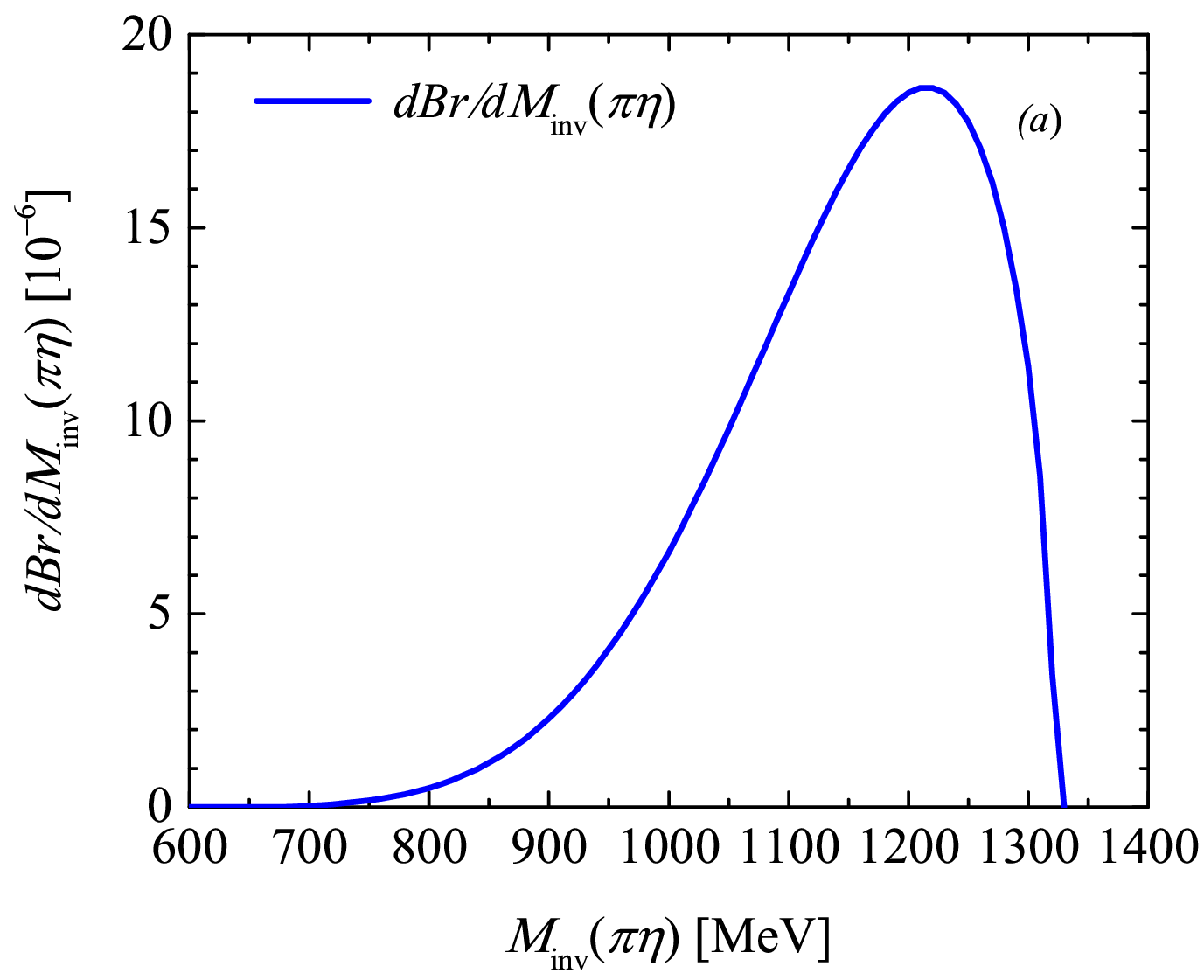 }
  \includegraphics[width=0.35\textwidth]{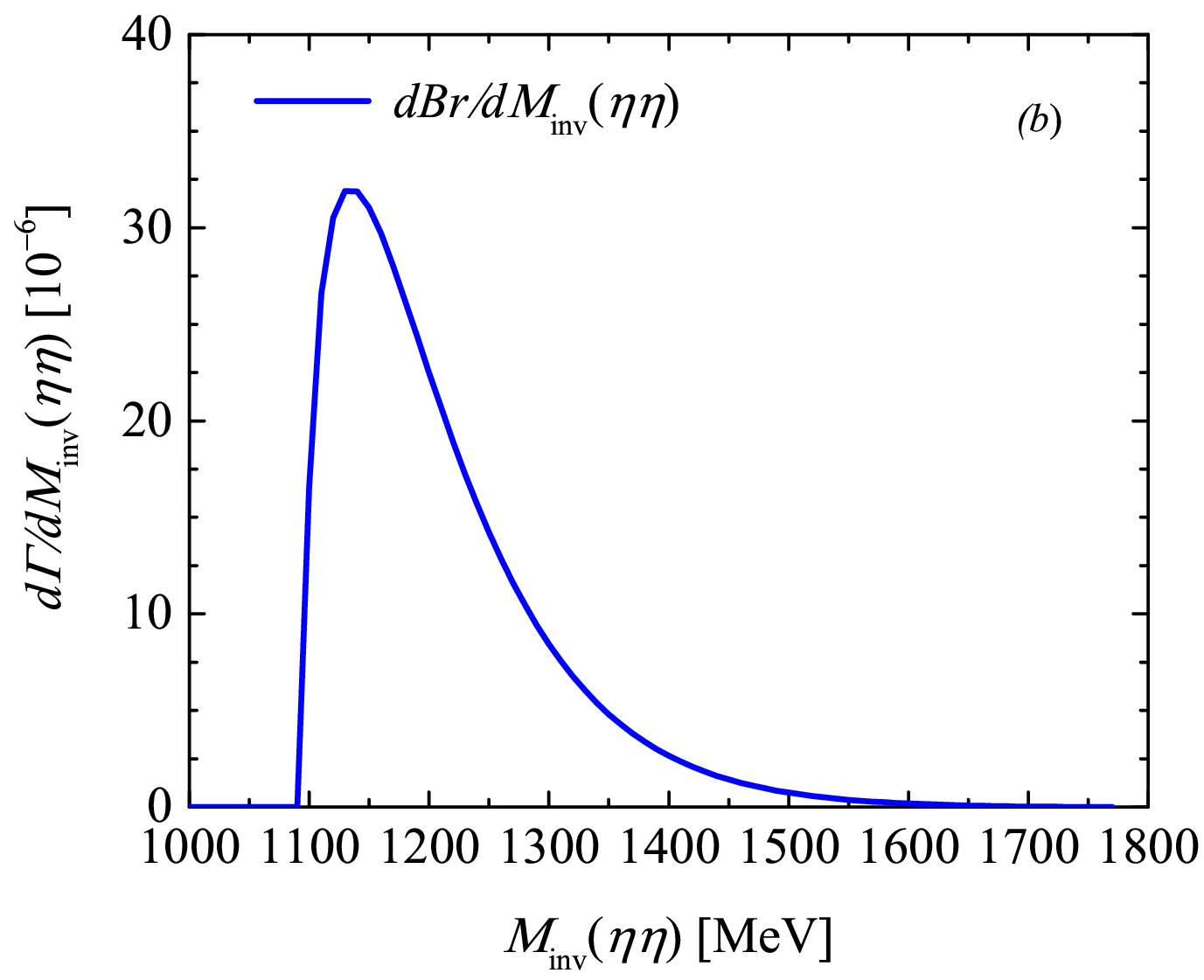 }
  \caption{ (a) $\pi^+\eta$ mass distribution, (b) $\eta\eta$ mass distribution considering only the $f_0(980)$ contribution. }
  \label{fig:RES_f980_individual}
\end{figure}

\begin{figure}[H]
  \centering
  \includegraphics[width=0.35\textwidth]{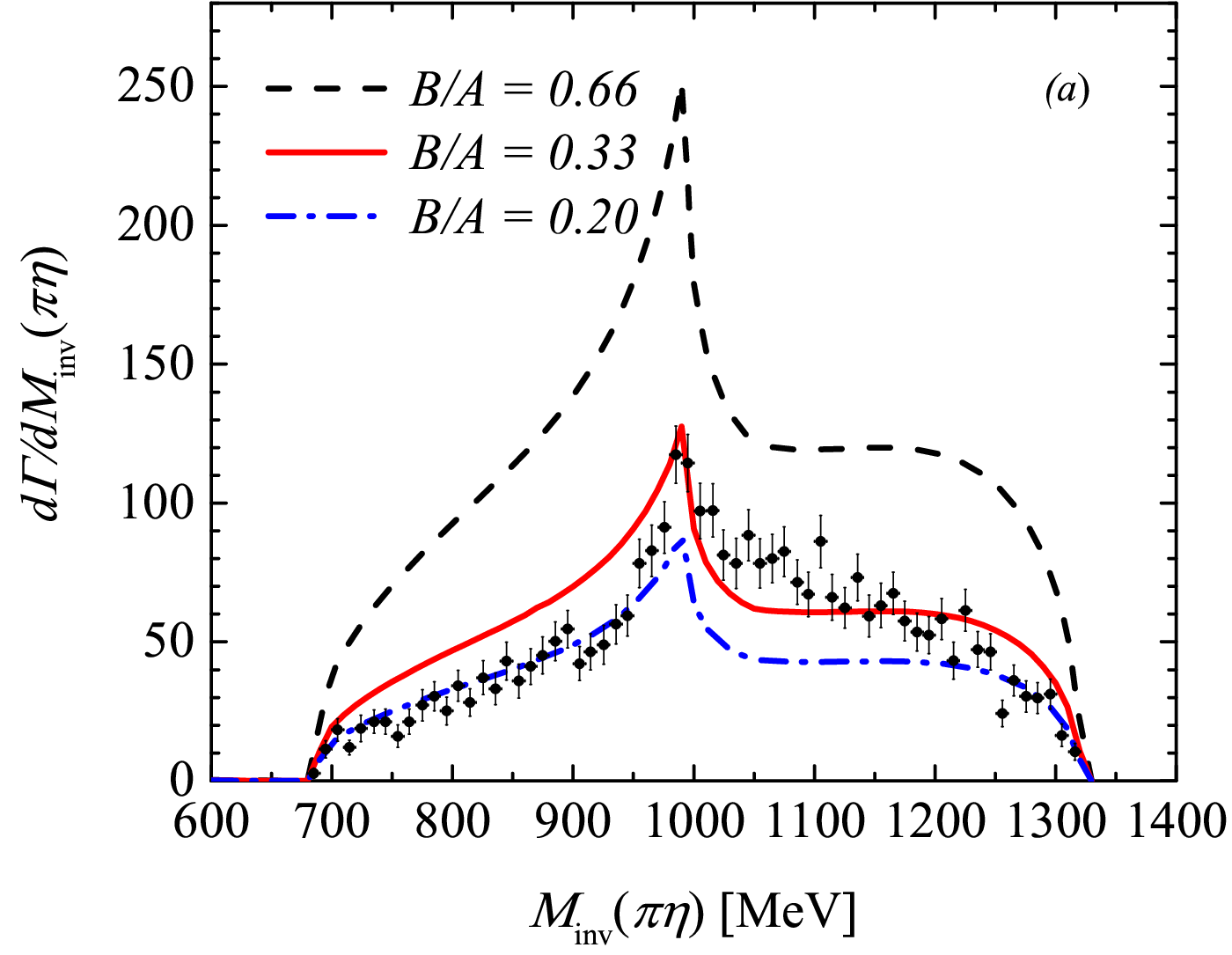 }
  \includegraphics[width=0.35\textwidth]{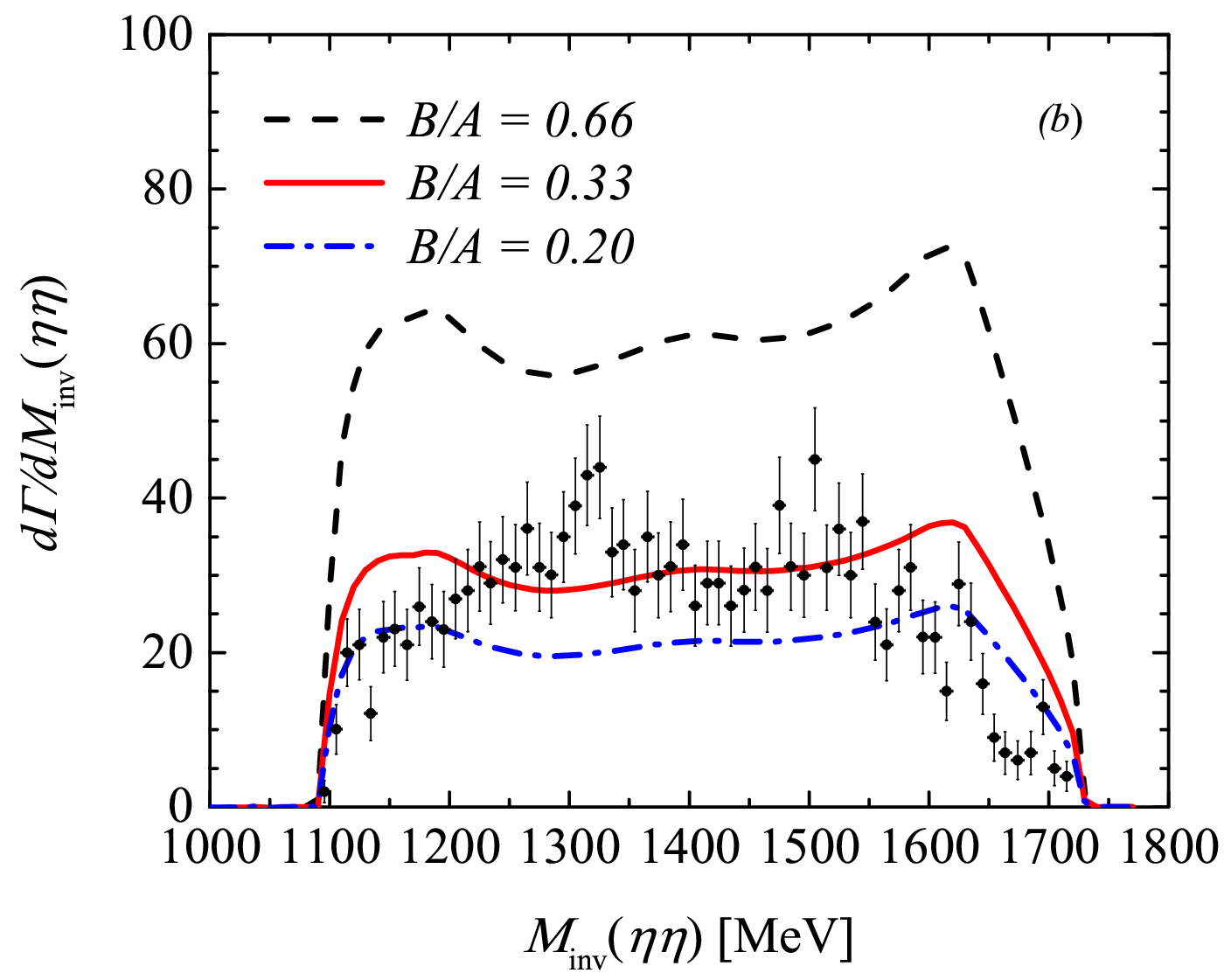 }
  \caption{(a) $M_{\pi^+\eta}$ and (b) $M_{\eta\eta}$ mass distributions obtained with $q_{\text{max}} = 650$ MeV, $M_{\text{cut}} = 1050$ MeV, $\alpha = 0.0060$ MeV$^{-1}$, and different $B/A$ ratios. The $y$ axis corresponds to events per 10 MeV of Ref.~\cite{BESIII:2025yag}.}
  \label{fig:a0+f980_Bplay}
\end{figure}

{
In Fig.~\ref{fig:RES_f980_individual}, we show the results for the $D^+ \to \pi^+ \eta \eta$ decay considering only the $t_{f_0}$ amplitude of Eq.~(\ref{eq:tf980}), which represents the contribution from the $f_0(980)$ resonance.
 As in Fig.~\ref{fig:RES_massdis}, the invariant mass in Figs. ~\ref{fig:RES_f980_individual} (a) and (b) are taken to be the same.
The calculations are done using the same values of $A$ and $B$ in Fig.~\ref{fig:RES_massdis}, $A=1,~ B=1/3$. This allows to compare the relative strengths of both mechanisms.
The first thing that we realize is that  the strength of the $f_0$ in the mass distributions is similar to that of the $a_0$.
As expected, the effect of the $f_0(980)$ is clearly seen in the $\eta\eta$ invariant mass distribution, producing a pronounced enhancement close to 1.0~–~1.1~GeV, which is most welcome in view of the experimental mass distribution. At the same time, it provides a sizeable contribution at large $\pi\eta$ invariant masses, again most welcome by the experiment.
}

In Fig.~\ref{fig:a0+f980_Bplay} we show the results adding $t_{a_0}$ and $t_{f_0}$ of  {Eqs. (\ref{eq:tRES}),  (\ref{eq:tf980}).} Here we need to differentiate between $A$ and $B$. Hence, apart from the absolute normalization of the mass distribution, we need the ratio $B/A$. This ratio should be of the order of $1/N_c (N_c = 3,\text{the number of the colors})$, but in order to make estimates of uncertainties we shall vary 
\begin{align}\label{ratioBA}
    B/A \in [0.2-0.66].
\end{align}
We have also tried with $B/A$ negative but the results are clearly bad, refusing this option.

First we show the results in Fig.~\ref{fig:a0+f980_Bplay}, taking as a reference $M_{\text{cut}} = 1050 $ MeV,  $\alpha = 0.0060 $ MeV$^{-1}$, taking different values of $B/A$.
In Fig.~\ref{fig:a0+f980_Bplay} we have already summed the mass distributions $\pi^+\eta(1)$ and $\pi^+\eta(2)$ as in the experiment~\cite{BESIII:2025yag}, such that the integrated $\pi^+\eta$ mass distribution has double strength than that of $\eta\eta$. The strength of $A$ is adjusted to have an integrated mass distribution equal to that of the counts in the experiment for $B/A=0.33$.

What we observe is that now we get a substantial strength in the $\pi^+\eta$ mass distribution at large $\pi^+\eta$ invariant masses, getting a shape roughly in agreement with experiment. On the other hand, the $\eta\eta$ mass distribution is also tremendously improved, obtaining now strength at lower $\eta\eta$ invariant masses that did not appear with the mechanism of the $a_0(980)$ alone.

\begin{figure}[H]
  \centering
  \includegraphics[width=0.35\textwidth]{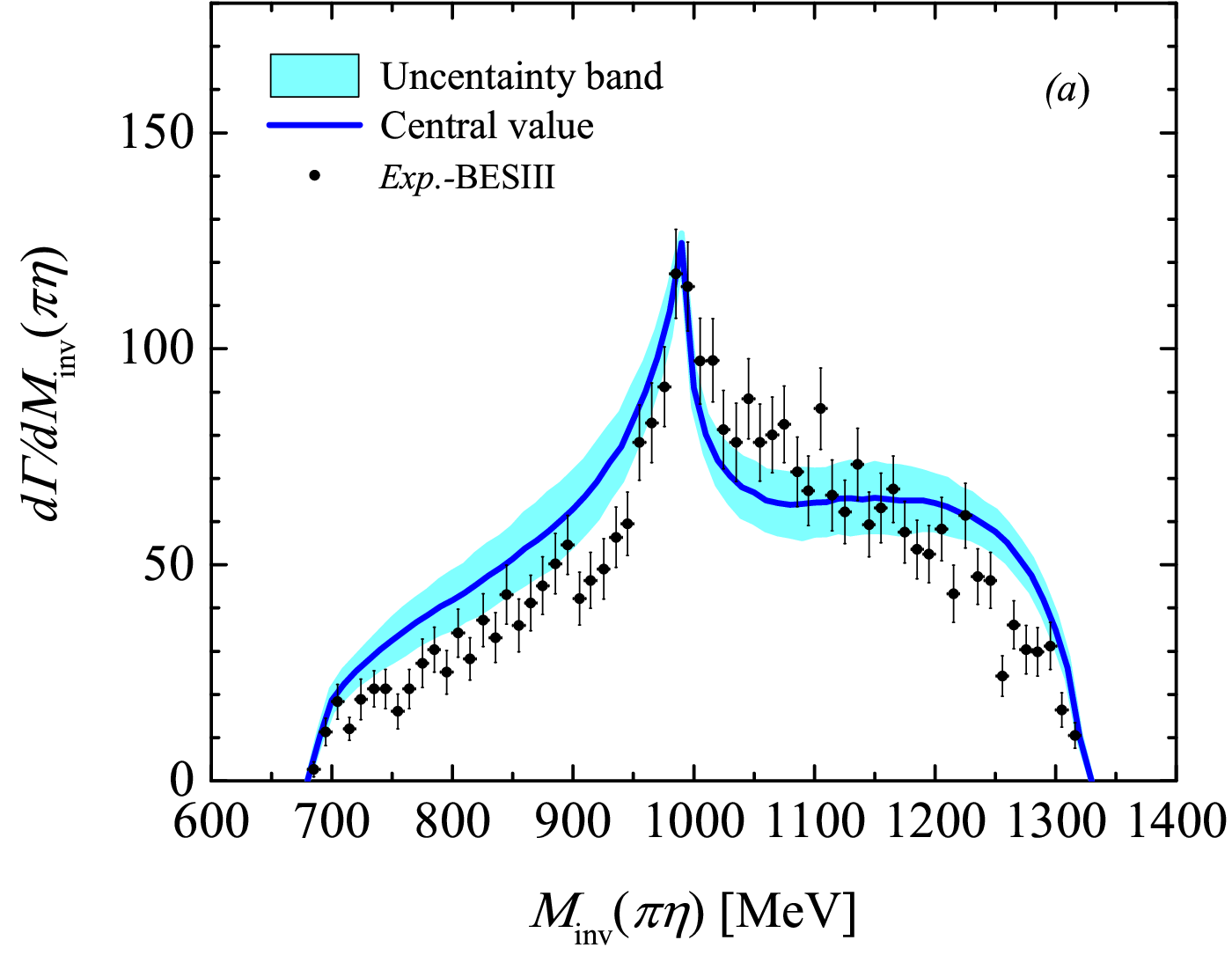 }
  \includegraphics[width=0.35\textwidth]{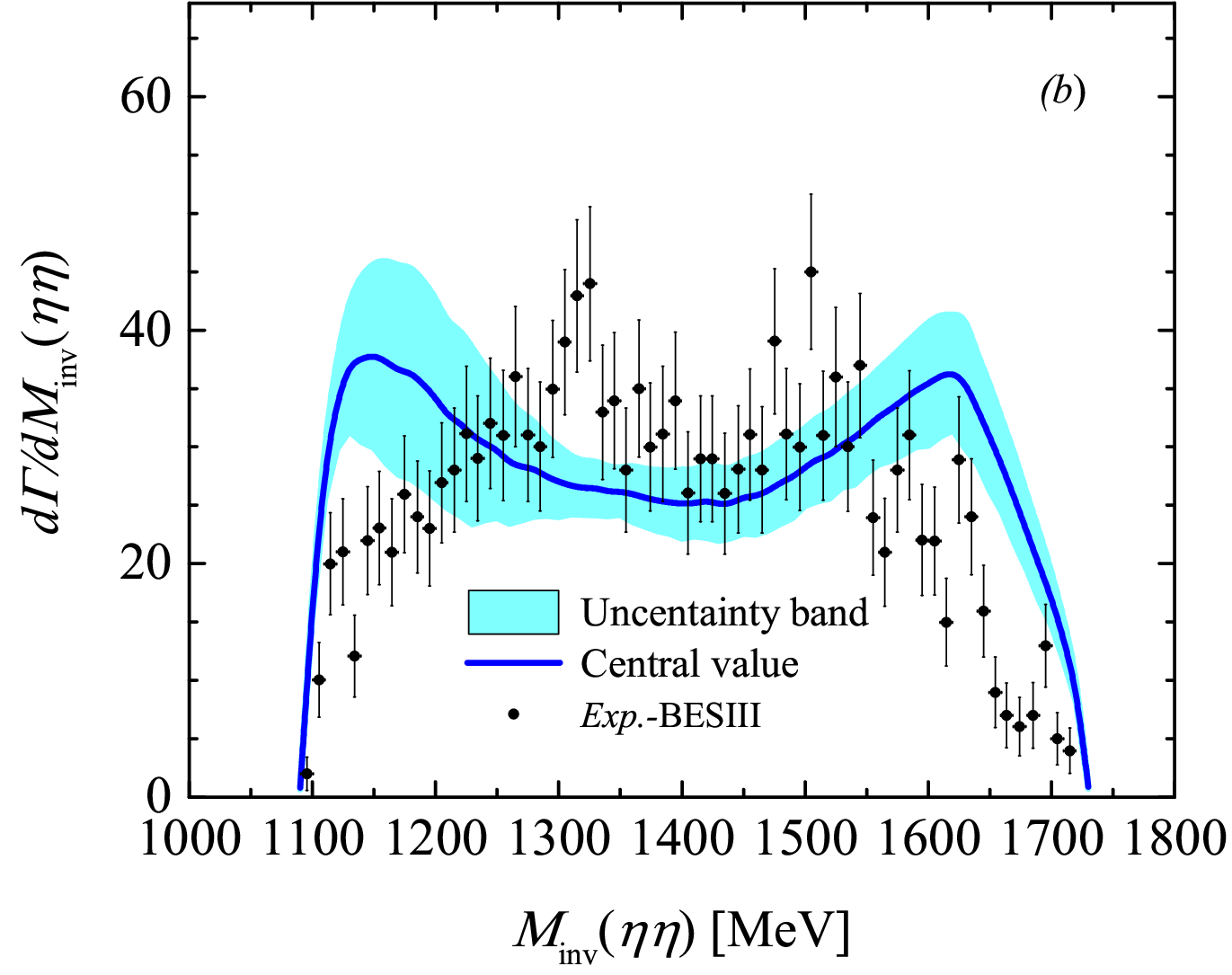 }
  \caption{Mass distributions of the $\pi \eta $ and $\eta \eta$ systems at $q_{\max }=650 \mathrm{MeV}$, showing the uncertainty band for $M_{\text {cut }}=1050-1150 \mathrm{MeV}$, $\alpha=0.0040-0.0080 \mathrm{MeV}^{-1}$, and $B / A=0.2-0.66$. The $y$ axis corresponds to events per 10 MeV of Ref.~\cite{BESIII:2025yag}. }
  \label{fig:massdisband}
\end{figure}

We should note that the two mechanisms of $a_0(980)$ and $f_0(980)$ production are linked in our approach. Up to an overall normalization, their relative strength is tied to the ratio of $B/A$, which should not be too far away from $1/3$. Then, Fig.~\ref{fig:a0+f980_Bplay} gives us an idea of the contribution that we have. In order to have a band of uncertainties of our results, we make a few runs of calculations choosing random numbers of $B/A$, $M_{\text{cut}}$, $\alpha$ in the range
\begin{align}\label{ratiopara}
    \alpha~~ &\in [0.0040-0.0080]~\text{MeV}^{-1},\\\nonumber
    M_{\text{cut}}~ &\in [1050-1150]~\text{MeV},\\\nonumber
    B/A~~ &\in [0.3-0.8].
\end{align}
The results are shown in Fig.~\ref{fig:massdisband}, but all of them have been normalized to the experimental integrated decay width.
We remove $16\%$ of the results in the lower and upper extremes to have $68\%$ confidence level.
Considering the theoretical error band, the agreement with the data can be considered fair, as good as one could expect within the approximations done. Nevertheless, the obvious improvements obtained when considering the $f_0(980)$ contribution tell us unequivocally that the $f_0(980)$ excitation with decay to $\eta\eta$ has been essential to reproduce the experimental features, in particular the large strength in the $\pi^+\eta$ mass distribution at large invariant masses, which make the shape of the $a_0(980)$ very different from that in other experiments as noted in~\cite{BESIII:2025yag}, and the large strength of the $\eta\eta$ mass distribution at low $\eta\eta$ invariant masses.

\section{Conclusions}
 We have made a thorough investigation of possible contributions to the $D^+ \to \pi^+ \eta \eta $ decay. The motivation of the work stems from the recent experiment in BESIII~\cite{BESIII:2025yag}, observing a $\pi^+ \eta$ mass distribution very different than that observed in most experiments producing the $a_0(980)$. The large strength in the higher mass region of the $\pi^+ \eta$ mass spectrum, prompted the authors of the BESIII team to suggest that it should be due to a mechanism of production based on a triangle diagram, induced by the $D^+$ decay into $K^*_0(1430)$ and $\bar K$, with subsequent decay of $K^*_0(1430)$  into $\eta K$ and final fusion of the $K \bar K$ to produce the $a_0(980)$, which decays to $\pi^+ \eta$.  In a following work  \cite{Lyu:2025oow}, the triangle mechanism was dismissed, because it does not develop a triangle singularity for the nominal mass of the $K^*_0(1430)$, and instead, the $D^+ \to \pi^+ f_0(1370)$; $f_0 \to \eta \eta$  mechanism was suggested to interpret the results. 
    We have done a thorough study of all these suggested solutions and concluded that the triangle mechanism develops, indeed, a triangle singularity, taking into account the large width of the $K^*_0(1430)$ state. However, we were able to make a determination of the width in absolute terms, not just the shape, and found that its contribution to the process is two orders of magnitude smaller than the experimental branching ratio.
Similarly, we also studied the $D^+ \to \pi^+ f_0(1370)$; $f_0 \to \eta \eta$  mechanism suggested in \cite{Lyu:2025oow} and could also determine its strength in absolute terms, finding also a contribution two orders of magnitude smaller than experiment. We also studied the $D^+ \to \eta a_0(1450)$; $a_0 \to \eta \pi^+$, and found cancellations that make null its contribution. Similarly, we also studied the possible contribution  of $D^+ \to \pi^+ f'_2(1525)$; $f'_2 \to \eta \eta$, and also found it extremely small and with small interference with the dominant decay mechanism, given the $D-$wave character of the $f'_2 \to \eta \eta$ decay. 

  Surprisingly, we found a natural solution to the problem by looking at the related, unavoidable contribution of the $f_0(980)$ excitation, which is linked to the $a_0(980)$ production. The $f_0(980)$ resonance, decaying to $\eta \eta$, is only operative through the tail of the resonance, but the weight of the amplitude is larger than that for $a_0(980)$ production and, at the end, the two mechanisms are competitive. The contribution of the $f_0(980)$ solves the problem of the large strength obtained at hight invariant masses in the $\pi^+ \eta$ mass distribution, and also fixes the problem of the small strength at low invariant masses of the $\eta \eta$ mass distribution, stemming from the $a_0(980)$ production.

\section*{ACKNOWLEDGMENTS}
We acknowledge useful discussions with Prof. Jia Jun Wu and Prof. En Wang.
This work is partly supported by the National Natural Science
Foundation of China under Grants  No. 12405089 and No. 12247108 and
the China Postdoctoral Science Foundation under Grant
No. 2022M720360 and No. 2022M720359.  Yi-Yao Li is supported in part by the Guangdong Provincial international exchange program for outstanding young talents of scientific research in 2024.
This work is partly supported by the Spanish Ministerio de Economia y Competitividad (MINECO) and European FEDER funds under Contracts No. FIS2017-84038-
C2-1-P B, PID2020-112777GB-I00, and by Generalitat
Valenciana under contract PROMETEO/2020/023. This
project has received funding from the European Union
Horizon 2020 research and innovation programme under
the program H2020-INFRAIA-2018-1, grant agreement
No. 824093 of the STRONG-2020 project.  This work
is partly supported by the Science and Technology Facilities Council (UK).

\bibliography{refs.bib} 
\end{document}